\documentclass[twocolumn,trackchanges]{aastex62}

\let\pwiflocal=\iffalse \let\pwifjournal=\iffalse
\usepackage[utf8]{inputenc}
\usepackage{textcomp}
\usepackage{amsmath,amssymb,gensymb,times,graphicx,morefloats,color}
\usepackage{afterpage, bm}
\usepackage{natbib,hyperref}
\usepackage{mathrsfs}

\newcommand{\vsini}{$v\sin{i_*}$}

\newcommand{\fbol}{$F_{\mathrm{bol}}$}

\newcommand{\teff}{\ensuremath{T_{\text{eff}}}}
\newcommand\kms{km~s$^{-1}$}

\def\spitzer{\emph{Spitzer}}
\def\Spitzer{\emph{Spitzer}}

\def\tess{\emph{TESS}}
\def\TESS{\emph{TESS}}

\newcommand{\ktwo}{\emph{K2}}
\newcommand{\system}{DS~Tuc}
\newcommand{\target}{DS~Tuc A}
\newcommand{\comp}{DS~Tuc B}

\shorttitle{A 45 Myr planet in Tuc-Hor}
\shortauthors{Newton et al. }

\begin{document}

\title{{\it TESS} Hunt for Young and Maturing Exoplanets (THYME):
A planet in the 45 Myr Tucana--Horologium 
association}
\correspondingauthor{Elisabeth R. Newton}
\email{Elisabeth.R.Newton@Dartmouth.edu}

\author[0000-0003-4150-841X]{Elisabeth R. Newton}
\affiliation{Department of Physics and Astronomy, Dartmouth College, Hanover, NH 03755, USA}
\affiliation{Department of Physics and Kavli Institute for Astrophysics and Space Research, Massachusetts Institute of Technology, Cambridge, MA 02139, USA}

\author[0000-0003-3654-1602]{Andrew W. Mann}
\affiliation{Department of Physics and Astronomy, The University of North Carolina at Chapel Hill, Chapel Hill, NC 27599, USA} 

\author[0000-0003-2053-0749]{Benjamin M. Tofflemire}
\affiliation{Department of Astronomy, The University of Texas at Austin, Austin, TX 78712, USA}

\author[0000-0003-3904-7378]{Logan Pearce}
\affiliation{Department of Astronomy, The University of Texas at Austin, Austin, TX 78712, USA}

\author[0000-0001-9982-1332]{Aaron C. Rizzuto}
\altaffiliation{51 Pegasi b Fellow}
\affiliation{Department of Astronomy, The University of Texas at Austin, Austin, TX 78712, USA}

\author[0000-0001-7246-5438]{Andrew Vanderburg}
\altaffiliation{NASA Sagan Fellow}
\affiliation{Department of Astronomy, The University of Texas at Austin, Austin, TX 78712, USA}

\author[0000-0001-6301-896X]{Raquel A. Martinez}
\affiliation{Department of Astronomy, The University of Texas at Austin, Austin, TX 78712, USA}

\author[0000-0003-0774-6502]{Jason J. Wang}
\altaffiliation{51 Pegasi b Fellow}
\affiliation{Department of Astronomy, California Institute of Technology, Pasadena, CA 91125, USA}

\author[0000-0003-2233-4821]{Jean-Baptiste Ruffio}
\affiliation{Kavli Institute for Particle Astrophysics and Cosmology, Stanford University, Stanford, CA, USA 94305}

\author[0000-0001-9811-568X]{Adam L. Kraus}
\affiliation{Department of Astronomy, The University of Texas at Austin, Austin, TX 78712, USA}

\author[0000-0002-5099-8185]{Marshall C. Johnson}
\affiliation{Department of Astronomy, The Ohio State University, Columbus, OH 43210, USA}

\author[0000-0001-5729-6576]{Pa Chia Thao}
\affiliation{Department of Physics and Astronomy, The University of North Carolina at Chapel Hill, Chapel Hill, NC 27599, USA} 

\author[0000-0001-7336-7725]{Mackenna L. Wood}
\affiliation{Department of Physics and Astronomy, The University of North Carolina at Chapel Hill, Chapel Hill, NC 27599, USA} 

\author[0000-0001-7337-5936]{Rayna Rampalli}
\affiliation{Department of Astronomy, Columbia University, 550 West 120th Street, New York, NY 10027, USA} 

\author[0000-0001-6975-9056]{Eric L. Nielsen}
\affiliation{Kavli Institute for Particle Astrophysics and Cosmology, Stanford University, Stanford, CA, USA 94305}


\author[0000-0001-6588-9574]{Karen A.\ Collins}
\affiliation{Center for Astrophysics Harvard and Smithsonian, 60 Garden St, Cambridge, MA, 02138, USA}

\author{Diana Dragomir}
\affiliation{Kavli Institute for Astrophysics and Space Science, Massachusetts Institute of Technology, Cambridge, MA 02139, USA}

\author{Coel Hellier}
\affiliation{Astrophysics Group, Keele University, Staffordshire ST5 5BG, U.K.}

\author{D. R. Anderson}
\affiliation{Astrophysics Group, Keele University, Staffordshire ST5 5BG, U.K.}

\author[0000-0001-7139-2724]{Thomas~Barclay}
\affiliation{NASA Goddard Space Flight Center, 8800 Greenbelt Road, Greenbelt, MD 20771, USA}
\affiliation{University of Maryland, Baltimore County, 1000 Hilltop Circle, Baltimore, MD 21250, USA}

\author{Carolyn Brown}
\affil{University of Southern Queensland, Centre for Astrophysics, West Street, Toowoomba, QLD 4350 Australia}

\author[0000-0002-2012-7215]{Gregory Feiden}
\affiliation{Department of Physics, University of North Georgia, Dahlonega, GA 30597, USA}

\author{Rhodes Hart}
\affiliation{Centre for Astrophysics, University of Southern Queensland, Toowoomba, QLD, 4350, Australia}

\author{Giovanni Isopi} 
\affiliation{Campo Catino Astronomical Observatory, Regione Lazio, Guarcino (FR), 03010 Italy}

\author[0000-0003-0497-2651]{John F.\ Kielkopf} 
\affiliation{Department of Physics and Astronomy, University of Louisville, Louisville, KY 40292, USA}

\author{Franco Mallia}
\affiliation{Campo Catino Astronomical Observatory, Regione Lazio, Guarcino (FR), 03010 Italy}

\author[0000-0001-7945-0634]{Peter Nelson}
\affiliation{Ellinbank Observatory, Australia}

\author[0000-0001-8812-0565]{Joseph E.\ Rodriguez}
\affiliation{Center for Astrophysics Harvard and Smithsonian, 60 Garden St, Cambridge, MA, 02138, USA}

\author[0000-0003-2163-1437]{Chris Stockdale}
\affiliation{Hazelwood Observatory, Australia}

\author[0000-0002-3249-3538]{Ian A. Waite}
\affiliation{Centre for Astrophysics, University of Southern Queensland, Toowoomba, QLD, 4350, Australia}

\author{Duncan J. Wright}
\affil{University of Southern Queensland, Centre for Astrophysics, West Street, Toowoomba, QLD 4350 Australia}

\author[0000-0001-6513-1659]{Jack Lissauer}
\affiliation{NASA Ames Research Center, Moffett Field, CA, 94035, USA}


\author[0000-0003-2058-6662]{George~R.~Ricker}
\affiliation{Department of Physics and Kavli Institute for Astrophysics and Space Research, Massachusetts Institute of Technology, Cambridge, MA 02139, USA}

\author[0000-0001-6763-6562]{Roland~Vanderspek}
\affiliation{Department of Physics and Kavli Institute for Astrophysics and Space Research, Massachusetts Institute of Technology, Cambridge, MA 02139, USA}

\author[0000-0001-9911-7388]{David~W.~Latham}
\affiliation{Center for Astrophysics Harvard and Smithsonian, 60 Garden St, Cambridge, MA, 02138, USA}

\author[0000-0002-6892-6948]{S.~Seager}
\affiliation{Department of Physics and Kavli Institute for Astrophysics and Space Research, Massachusetts Institute of Technology, Cambridge, MA 02139, USA}
\affiliation{Department of Earth, Atmospheric and Planetary Sciences, Massachusetts Institute of Technology, Cambridge, MA 02139, USA}
\affiliation{Department of Aeronautics and Astronautics, MIT, 77 Massachusetts Avenue, Cambridge, MA 02139, USA}

\author[0000-0002-4265-047X]{Joshua~N.~Winn}
\affiliation{Department of Astrophysical Sciences, Princeton University, 4 Ivy Lane, Princeton, NJ 08544, USA}

\author{Jon~M.~Jenkins}
\affiliation{NASA Ames Research Center, Moffett Field, CA, 94035, USA}

\author[0000-0002-0514-5538]{Luke~G.~Bouma}
\affiliation{Department of Astrophysical Sciences, Princeton University, 4 Ivy Lane, Princeton, NJ 08544, USA}

\author[0000-0002-7754-9486]{Christopher~J.~Burke}
\affiliation{Department of Physics and Kavli Institute for Astrophysics and Space Research, Massachusetts Institute of Technology, Cambridge, MA 02139, USA}

\author{Misty Davies}
\affiliation{NASA Ames Research Center, Moffett Field, CA, 94035, USA}

\author[0000-0002-9113-7162]{Michael~Fausnaugh}
\affiliation{Department of Physics and Kavli Institute for Astrophysics and Space Research, Massachusetts Institute of Technology, Cambridge, MA 02139, USA}

\author{Jie Li}
\affiliation{NASA Ames Research Center, Moffett Field, CA, 94035, USA}
\affiliation{SETI Institute, Mountain View, CA 94043, USA}

\author{Robert L. Morris}
\affiliation{NASA Ames Research Center, Moffett Field, CA, 94035, USA}
\affiliation{SETI Institute, Mountain View, CA 94043, USA}

\author{Koji Mukai} \affiliation{NASA Goddard Space Flight Center, 8800 Greenbelt Road, Greenbelt, MD 20771, USA}
\affiliation{University of Maryland, Baltimore County, 1000 Hilltop Circle, Baltimore, MD 21250, USA}

\author{Joel Villase{\~ n}or}
\affiliation{Department of Physics and Kavli Institute for Astrophysics and Space Research, Massachusetts Institute of Technology, Cambridge, MA 02139, USA}

\author{Steven Villeneuva} 
\affiliation{Department of Physics and Kavli Institute for Astrophysics and Space Research, Massachusetts Institute of Technology, Cambridge, MA 02139, USA}


\author[0000-0002-4918-0247]{Robert J. De Rosa}
\affiliation{Kavli Institute for Particle Astrophysics and Cosmology, Stanford University, Stanford, CA, USA 94305}

\author[0000-0003-1212-7538]{Bruce Macintosh}
\affiliation{Kavli Institute for Particle Astrophysics and Cosmology, Stanford University, Stanford, CA, USA 94305}


\author{Matthew W. Mengel}
\affil{University of Southern Queensland, Centre for Astrophysics, West Street, Toowoomba, QLD 4350 Australia}

\author{Jack Okumura}
\affil{University of Southern Queensland, Centre for Astrophysics, West Street, Toowoomba, QLD 4350 Australia}

\author{Robert A. Wittenmyer}
\affil{University of Southern Queensland, Centre for Astrophysics, West Street, Toowoomba, QLD 4350 Australia}

\begin{abstract}

    Young exoplanets are snapshots of the planetary evolution process. Planets that orbit stars in young associations are particularly important because the age of the planetary system is well constrained. We present the discovery of a transiting planet larger than Neptune but smaller than Saturn in the 45\,Myr Tucana--Horologium young moving group. The host star is a visual binary, and our follow-up observations demonstrate that the planet orbits the G6V primary component, DS Tuc A (HD 222259A, TIC 410214986). 
    We first identified transits using photometry from the {\it Transiting Exoplanet Survey Satellite} (\tess; alerted as TOI 200.01). We validated the planet and improved the stellar parameters using a suite of new and archival data, including spectra from SOAR/Goodman, SALT/HRS and LCO/NRES; transit photometry from {\it Spitzer}; and deep adaptive optics imaging from Gemini/GPI.  No additional stellar or planetary signals are seen in the data. We measured the planetary parameters by simultaneously modeling the photometry with a transit model and a Gaussian process to account for stellar variability. We determined that the planetary radius is $5.70\pm0.17$ $R_\earth$ and that the orbital period is $8.1$ days.
    The inclination angles of the host star's spin axis, the planet's orbital axis, and the visual binary's orbital axis are aligned within $15$\degree\ to within the uncertainties of the relevant data. DS~Tuc~Ab is bright enough ($V=8.5$) for detailed characterization using radial velocities and transmission spectroscopy.
    
\end{abstract}



\section{Introduction}

 Exoplanets do not form with the properties with which we observe them today: migration and dynamical interactions change their orbital parameters, high-energy radiation from their host stars causes atmospheric mass loss, and gaseous planets contract as they cool. The demographics of field-age (typically $>1$ Gyr) exoplanetary systems offers one way to learn about the evolutionary history of exoplanets. For example, the gap in the observed radius distribution of close-in planets (between super-Earths and mini-Neptunes) has been used as a probe of photoevaporation and to constrain typical core compositions \citep{OwenEvaporation2017, LopezBorn2017}; and \citet{Owen2018} explained the dearth of close-in giant planets  as the joint result of high-eccentricity migration and photoevaporation. 
 
Observations of planets young enough to still be undergoing dynamical and atmospheric changes provide a more direct way to probe planetary evolution; and planets in young stellar associations are particularly useful because the ages of these systems are known more precisely and accurately than those of their counterparts in the galactic field. The typically close-orbiting planets discovered through transit and radial velocity surveys complement the constraints on planet formation beyond the snow line available from direct imaging \citep[e.g.][]{2014ApJ...794..159B, 2015ApJS..216....7B, 2016ApJ...819..125C, NielsenGPIES}. They are also likely to be young representatives of the field-age exoplanets on which planetary demographics studies are based.

Radial velocity programs have detected Jupiter mass planets in young clusters \citep{QuinnTwo2012, QuinnHD2014}, but are hindered by the radial velocity jitter exhibited by these young, active stars \citep[e.g.][]{SaarActivityRelated1997, 2004AJ....127.3579P}. Thanks to its excellent photometric precision and wide-area coverage, \ktwo\ yielded a surge of exoplanet discoveries around young stars via the transit method. This included planets in the Hyades \citep{MannZodiacal2016a, DavidNew2016}, Upper Scorpius \citep{DavidNeptunesized2016, MannZodiacal2016}, Praesepe \citep{MannZodiacal2017, RizzutoZodiacal2018, LivingstonK22642019}, and Taurus-Auriga \citep{DavidWarm2019} associations. 
 
The {\it Transiting Exoplanet Survey Satellite} (\tess) will survey 80\% of the sky during its prime mission, with a focus on bright stars. \tess\ enables the transit search for young exoplanets in associations to be substantially expanded; and motivates our collaboration, the \tess\ Hunt for Young and Maturing Exoplanets (THYME) Project. 
 
\tess\ provides the first opportunity for extensive transit surveys of stars in young moving groups (YMGs). YMGs are dynamically unbound associations of stars that are identified based on their common motion through the galaxy. YMGs have ages $\lesssim300$ Myr; and probe a more continuous range of ages than do young stellar clusters \citep[see e.g.][]{Bell2015}. The stellar environments in YMGs also differ from those found in high-density, longer-lasting star clusters such as Praesepe or Pleiades.  These clusters are less compact and therefore stellar dynamical interactions are less frequent; as a result, they may be more characteristic of the precursors of exoplanetary systems that orbit typical field stars. Dynamical studies indicate that stellar interactions in open clusters are unlikely to disrupt planetary systems \citep[e.g][]{2001MNRAS.322..859B, 2006ApJ...641..504A}, but milder impacts, such as changes in eccentricity, are possible \citep{2009ApJ...697..458S}.
Finally, most known YMGs are substantially less distant than stellar clusters \citep[see e.g.][]{Gagne2018}. This provides significant advantages for detailed characterization of the planets through techniques such as transmission spectroscopy and precise radial velocity monitoring.
 
\begin{figure*}
    \centering
    \includegraphics[width=0.8\textwidth]{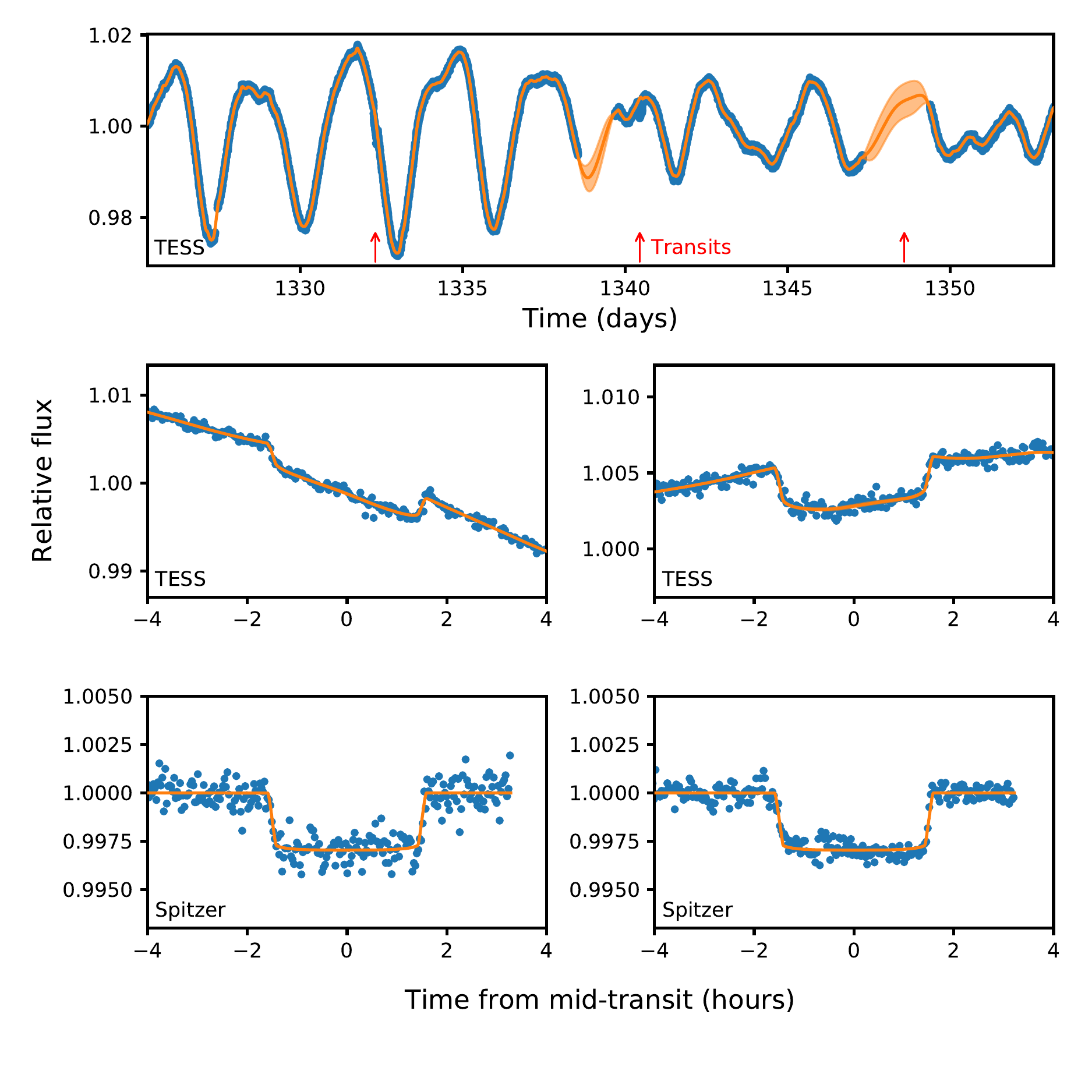}\vspace{-0.8cm}
    \caption{Discovery data from \tess, after our iterative flare rejection algorithm has been applied, and follow-up data from \spitzer. Data are shown as blue points; data for \spitzer\ are the means of 250 equally spaced bins. The top panel shows the full \tess\ lightcurve and the stellar variability Gaussian process (GP) model. The middle panel shows a zoom-in on the two transits observed with \tess. The bottom panel shows the two \spitzer\ transits at $4.5\micron$. The best fitting model from our joint fit to the these lightcurves is shown in orange; in this analysis we simultaneously model stellar variability in \tess, using a GP, and the transit parameters.  The mean of the MCMC samples is shown as the opaque orange line; the $1\sigma$ deviations are shown as the semi-transparent orange region. }
    \label{fig:transit}
\end{figure*}

We report the discovery (Figure \ref{fig:transit}) of a close-in, transiting planet with a radius in between those of Neptune and Saturn. The stellar host is the primary component of DS Tuc (\target, HD~222259A), which is a member of the Tucana--Horologium (Tuc-Hor) YMG. DS Tuc was one of the original members of the Tucana association of co-moving stars identified by \cite{ZuckermanIdentification2000}. Tucana was soon identified as being physically associated with the Horologium association of active stars \citep{TorresNew2000}, and together they formed one of the first known YMGs. 
 
DS Tuc is a visual binary \citep[]{TorresVisual1988}, consisting of a G6V primary and a K3V secondary \citep{TorresSearch2006} separated by $5\arcsec$. \citet{SoderblomHighResolution1998} suggested that the secondary (\comp, HD~222259B) is itself a short period binary based on radial velocity variations, and \cite{CutispotoFastrotating2002} report spectral types for the components of K3/4V and K5V but do not provide further information. As we will discuss in Section~\ref{sec:rvs}, our radial velocity measurements demonstrate that \comp\ is not likely to be a short-period binary.
 
In Section \ref{sec:data} we present discovery data from \tess\ and follow-up photometry from {\it Spitzer}. We additionally present new high resolution spectra and long-term photometric monitoring, and discuss archival high resolution spectra. In Section \ref{Sec:measurements} we update the stellar parameters, and analyze the radial velocities and stellar rotation. In Section \ref{sec:system}, we investigate the overall DS Tuc system, including modeling of the binary star orbit, and a searching for additional companions in high contrast imaging and in the \tess\ transit data. We present the results of our transit analysis, including identifying the stellar host as DS Tuc A and assessing false-positive scenarios, in Section \ref{sec:planet}. We discuss the overall system architecture and prospects for future follow-up in Section \ref{Sec:discussion} and briefly summarize our findings in Section \ref{Sec:summary}.
 
\section{Observations}\label{sec:data}

\subsection{Photometry}

\subsubsection{\tess}

\tess\ was launched on 2018 April 18 and commenced science operations on 2018 July 25. \tess\ uses its four small (10 cm effective aperture) cameras to monitor 24\degree $\times$96\degree\ sectors of sky nearly continuously over $27$ day campaigns. DS Tuc was observed in the first sector of science operations during late July and August of 2018 and was pre-selected for fast (two-minute) cadence observations because of its membership in the young Tucana--Horologium Moving Group.\footnote{The target was requested as part of our Guest Investigator program GO11175 (PI: Mann), as well as by GO11176 (PI: Czekala) and GO11250 (PI: Walter)} After the \tess\ data were downlinked to Earth, they were processed by the Science Processing and Operations Center (SPOC) pipeline at NASA Ames \citep{Jenkins:2015, Jenkins:2016}, which calibrated the \tess\ pixels, extracted light curves, de-blended light from nearby contaminating stars, removed common-mode systematic errors, high-pass filtered the light curve, and searched for transits. We used the pre-search data condition simple aperture photometry (PDC-SAP) light curve and systematics solution throughout this paper, masking the time $1346.5<t<1350$, except in our transit injection and recovery tests (Section~\ref{sec:notchinjrec}). This time corresponds to the loss of fine guiding, where $t$ is given in \tess\ barycentric Julian date (BJD$-2457000.0$).

SPOC used the Transiting Planet Search module (TPS) to search for transits in the PDC-SAP data, applying a matched filter to compensate for stellar variability. TPS identified several ``threshold crossing events,'' or possible transiting planet signals (TCEs), in the light curves of both DS Tuc A and B. Upon visual inspection of results from the initial run of TPS, our team of vetters concluded that while the periodicities detected by TPS did not corresponded to transiting planets, some of the TCEs appeared transit-like. We identified two candidate transits 8.1 days apart; a third happened to fall during the three day period of time when \tess\ lost fine guiding. We alerted the community to the detection via the MIT \tess\ Alerts webpage\footnote{\url{https://tess.mit.edu/alerts/}} under the designation TOI-200.
We note that the alert was issued in early November based on the first TPS run from late August.  The second, archival TPS run from mid September, which was not included in the alert, detected a TCE that corresponds to DS Tuc Ab and that passed all diagnostic tests in the data validation report.

\subsubsection{\spitzer}\label{sec:spitzer}

Based on the \tess\ alert, we scheduled observations of two transits with the \spitzer\ Space Telescope, which were conducted on 2019 March 01 and 2019 March 09 UTC (Program ID: 14011, PI: Newton). We observed at $4.5\micron$ (channel 2) using the Infrared Array Camera \citep[IRAC; ][]{FazioInfrared2004}. We used the 32$\times$32 pixel subarray, and due to the brightness of DS Tuc A, we used $0.4$ second frame times. We followed the suggestions of \citet{IngallsIntrapixel2012,IngallsRepeatability2016}, placing DS Tuc A in the ``sweet spot'' of the detector and using the ``peak-up'' pointing mode to keep the position of the star fixed to within a half-pixel. Each transit observation consisted of a 30 minute dither, a $7.5$ hour stare including the full transit, and a final 10 minute dither.
Both \target\ and B are present in the \spitzer\ images. In the post-cryogenic mission, IRAC has a pixel scale of $1.2\arcsec$/pixel and a full-width at half-maximum of $2.0\arcsec$, so the binary components are resolved but not well-separated ($4.5$ pixels). 

To address the potential for flux dilution, we modeled the point spread functions (PSFs) of both components. 
We generated IRAC PSFs using the {\tt prf\_realize} routine as implemented in the software package {\tt IRACSIM}\footnote{\href{https://github.com/ingalls91104/IRACSIM}{https://github.com/ingalls91104/IRACSIM}} \citep{IngallsRepeatability2016} and incorporated them into the PSF-fitting framework described by Martinez \& Kraus (submitted to AAS Journals), modified for use with subarray images. To briefly summarize, we fit a two-source PSF model in each subarray image by performing an MCMC analysis using a standard Metropolis-Hastings algorithm with Gibbs sampling. The PSF model is described by seven parameters: $x$-pixel coordinate of the primary centroid ($x$), $y$-pixel coordinate of the primary centroid ($y$), image background ($b$), primary peak pixel value ($n$), projected separation ($\rho$), position angle (PA), and contrast ($\Delta m$). We ran four MCMC chains with 140,000 steps each, discarding the first $10$\% of each chain (the ``burn-in'' phase). Using the weighted average of the median ($x$,$y$)-centroid, $\rho$, PA, and $\Delta m$ generated by our MCMC fits, we made a single PSF model template of \comp. This method yielded an estimate for pixel-by-pixel flux contamination levels, which we use to select the best aperture. Based on this, we selected a fixed aperture of 4$\times$4 pixels, which minimized the level of contamination flux from \comp\ (2.2\%), while capturing $>$90\% of the flux from \target. 

Due to \spitzer's large intra-pixel sensitivity variations and its pointing jitter, the measured flux of the target can vary with time as the location of the star shifts on the detector \citep{IngallsIntrapixel2012}. To correct for this, we used a high-resolution pixel-sensitivity variation map \citep[PMAP,][]{IngallsIntrapixel2012}, following the recommendations from the IRAC website\footnote{\href{https://irachpp.spitzer.caltech.edu/page/contrib}{https://irachpp.spitzer.caltech.edu/page/contrib}} to calculate \target's centroid position and total flux in each image within the aperture given above. We then used the \texttt{iracpc\_pmap\_corr} routine to calculate corrected flux values. Further details about the photometric gain map are discussed by \citet{IngallsIntrapixel2012}.

\subsubsection{WASP}

DS Tuc was observed by the WASP-South station of the Wide Angle Search for Planets \citep[WASP;][]{PollaccoWASP2006} located in Sutherland, South Africa.  WASP-South consists of eight cameras on an equatorial mount, each with a 2048$\times$2048 CCD. Observations in 2010 and 2011 used 200\,mm, f/1.8 lenses with a broadband filter spanning $400-700$\,nm and a plate scale of $13.7\arcsec$/pixel. Observations from 2012 to 2014 used 85\,mm, f/1.2 lenses with a Sloan r' filter and a plate scale of $32\arcsec$/pixel.  

Approximately $74000$ observations of the DS Tuc system were obtained over $900$ nights spanning five years. DS Tuc A and B are not resolved in the WASP data, and the precision is not sufficient to detect the transit of DS~Tuc~Ab; these data are used to investigate the stellar rotation period (Section~\ref{sec:rotation}).

\subsection{Spectroscopy}

\subsubsection{SOAR/Goodman}\label{sec:goodman}

On 2018 December 23 we acquired moderate resolution spectra of both \target\ and \comp\ using the Goodman High Throughput Spectrograph \citep{Goodman} at the 4.1 m Southern Astrophysical Research (SOAR) Telescope located at Cerro Pach\'{o}n, Chile. We observed both targets at low airmass ($ \sec(z) \simeq 1.4$) with clear sky conditions using the 0.46\arcsec-long slit, 400\,lines/mm grating and M2 setup. This yielded moderate resolution ($R\simeq1850$) spectra spanning $5000-9000$\,\AA. 

After basic image reduction including bias and dark subtraction, and flat-fielding, we removed sky lines in the 2D image using the chip regions adjacent to the science spectrum in the spatial direction and cosmic rays by median stacking over 5 images of each target. We then optimally extracted the spectrum \citep{Horne1986} and applied a wavelength solution derived from HgAr lamp exposures taken just before the target observations. Lastly, we flux calibrated each spectrum using spectrophotometric standards taken during the night. These data are used to determine the stellar parameters (Section~\ref{Sec:stellarparams}).

\subsubsection{Archival data from HARPS, UVES, and FEROS}

We gathered processed archival spectra from HARPS, UVES, and FEROS using the ESO archive. While the FEROS spectrum is labeled as \comp\ in the ESO archive, the spectral features (in particular, the strength of $H\alpha$ and $H\beta$) clearly reveals that this spectrum belongs to \target. 
These data are used in our radial velocity analysis (Section~\ref{sec:rvs}).

\subsubsection{SALT/HRS}

We observed independent spectra of \target\ and \comp\ using the High Resolution Spectrograph \citep[HRS;][]{CrausePerformance2014} on the South African Extremely Large Telescope \citep[SALT;][]{2006SPIE.6267E..0ZB}. We obtained spectra on the nights of 2018 November 16, 18, 19, and 21. We used the high resolution mode, and spectra were reduced using the MIDAS pipeline.\citep{KniazevMN482016, KniazevSALT2017}\footnote{\href{http://www.saao.ac.za/~akniazev/pub/HRS_MIDAS/HRS_pipeline.pdf}{http://www.saao.ac.za/\texttildelow akniazev/pub/HRS\_MIDAS/HRS\_pipeline.pdf}} The pipeline performed flat fielding and wavelength calibration using ThAr and Ar lamps; we did not use the sky-subtracted or merged data. The nominal spectral resolutions of the blue and red arms are $65000$ and $74000$, respectively; however, the resolution achieved by the MIDAS pipeline is approximately $46000$ as a result of not accounting for the tilt of the spectral lines. 
These data are used in our radial velocity analysis (Section~\ref{sec:rvs}).

\subsubsection{NRES/LCO}

We observed one spectrum of DS Tuc A using Las Cumbres Observatory's \citep[LCO,][]{LCO2013} Network of Robotic Echelle Spectrographs \citep[NRES,][]{NRES2018} on UT 2018 December 11. Data were reduced automatically by the LCO NRES pipeline version 0.8\footnote{\href{https://github.com/LCOGT/nres-pipe}{https://github.com/LCOGT/nres-pipe}}, which included basic bias/dark corrections, optimal extraction of the one-dimensional spectrum, and wavelength calibration with ThAr lamps. The NRES pipeline also yielded a radial velocity estimate, but we used our own determination for consistency with other analyses (see Section~\ref{sec:rvs}). The final reduced spectra have a resolution of approximately $R\simeq53,000$ and cover 3800--8600\,\AA. The spectrum had SNR$>$50 per resolving element around the Mg b lines ($\simeq$5160\,\AA). These data are used in our radial velocity analysis (Section~\ref{sec:rvs}).

\subsection{High contrast imaging}
We performed $H$-band integral field spectroscopy of both stars using the Gemini Planet Imager \citep[GPI;][]{Macintosh2014}. As part of the GPI Exoplanet Survey (GPIES), \comp\ was observed on 2016 November 18 (program code GS-2015B-Q-500) and \target\ was observed on 2016 October 22 (GS-2015B-Q500) under poor conditions, aborted after 9 images, and then observed again under better conditions on 2016 November 18 (GS-2015B-Q-500). A high-order adaptive optics system compensated for atmospheric turbulence, and an apodized Lyot coronagraph was used to suppress starlight. Using 59.6~s integration times, we obtained 37.78~minutes of data with 14.9$^\circ$ of parallactic angle rotation for \comp\ and 4.97-minutes and 35.79~minutes of data with $5.0^\circ$ and $15.2^\circ$ of parallactic angle rotation for the two observations of \target. 

All three datasets were reduced using the GPIES automated data reduction pipeline \citep{Wang2018}. Briefly, the data were dark subtracted, a bad-pixel correction was applied, the microspectra positions determined using an Argon arc lamp snapshot taken right before each sequence, 3D spectral datacubes were extracted using wavelength solutions derived from deep Argon arc lamp data, the images were distortion corrected, and fiducial diffraction spots (satellite spots) were used to locate the position of the star in each image. The stellar point spread function (PSF) was then subtracted from each image using both angular differential imaging \citep{Marois2006} and spectral differential imaging \citep{Sparks2002} to disentangle the stellar PSF from any potential companions, and principal component analysis to model the stellar PSF \citep{Soummer2012,Wang2015}. The resulting image was then used to search for point sources (Section \ref{sec:directimaging}).

\subsection{Literature photometry \& astrometry} \label{sec:phot}

To better characterize the properties of each component we drew resolved photometry and astrometry for \target\ and \comp\ from the literature. Specifically, we adopted optical $B_T$ and $V_T$ photometry from the Tycho-2 Survey \citep{Hog2000}, optical $G$, $BP$, and $RP$ photometry from the second {\it Gaia} data release \citep[DR2;][]{Evans2018}, near-infrared $J$, $H$, and $K_S$ photometry from The Two Micron All Sky Survey \citep[2MASS,][]{Skrutskie2006}, and mid-infrared $W1$, $W2$, $W3$, and $W4$ photometry from the {\it Wide-field Infrared Survey Explorer} \citep[WISE;][]{Wright2010}. We also adopted proper motions and parallaxes for each component from DR2 \citep{GaiaDr2}, and J2000 positions from Tycho-2. 

All photometry and astrometry from the literature used in our analysis is listed in Table~\ref{tab:sparams}.

\floattable
\begin{deluxetable}{l c c l }
\tabletypesize{\footnotesize}
\tablecaption{Parameters of \system\ \label{tab:sparams}}
\tablewidth{0pt}
\tablehead{
\colhead{Parameter} & \colhead{DS Tuc~A}& \colhead{DS Tuc~B} & \colhead{Source}
}
\startdata
\multicolumn{4}{c}{{\bf Identifiers}} \\
TOI &\multicolumn{2}{c}{200.01}&  \\
Gaia DR2 &6387058411482257536 & 6387058411482257280& Gaia DR2 \\
TIC & 410214986 & 410214984 & \citet{TIC2018} \\
2MASS & J23393949-6911448 & J23393929-6911396& 2MASS \\
HD & 222259A & 222259B & \citet{Cannon1924_draper} \\
\hline
\multicolumn{4}{c}{{\bf Astrometry}} \\
$\alpha$ R.A. (hh:mm:ss J2000) & \phantom{$-$}23:39:39.49  &  \phantom{$-$} 23:39:39.27	 & Tycho-2\\
$\delta$ Dec. (dd:mm:ss J2000) & $-$69:11:44.88 & $-$69:11:39.51 & Tycho-2\\
$\mu_{\alpha}$ (mas~yr$^{-1}$) & 79.464$\pm$0.074 & 78.022$\pm$0.064  & Gaia DR2 \\
$\mu{\delta}$ (mas~yr$^{-1}$) &  -67.440 	$\pm$0.045  & -65.746 $\pm$ 0.037 & Gaia DR2 \\
$\pi$ (mas) & 22.666 $\pm$ 0.035 & 22.650 $\pm$ 0.030& Gaia DR2\\
\hline
\multicolumn {4}{c}{{\bf Photometry}} \\
$B_T$ (mag) & 9.320 $\pm$	0.017 & 10.921 $\pm$ 0.060 &Tycho-2 \\
$V_T$ (mag) & 8.548 $\pm$	0.012 & 9.653$\pm$ 0.030 &Tycho-2 \\
$G$ (mag) & 8.3193$\pm$0.0010 & 9.3993 $\pm$0.0014 &Gaia DR2 \\
$G_{BP}$ (mag) & 8.7044$\pm$0.0049 	 &9.9851$\pm$0.0059 &Gaia DR2 \\
$G_{RP}$ (mag) & 7.8137$\pm$0.0036 	 & 8.7082$\pm$0.0044&Gaia DR2 \\
$J$ (mag) & 7.122 $\pm$ 0.024 & 7.630 $\pm$ 0.058 &2MASS\\
$H$ (mag) & 6.759 $\pm$ 0.023 & 7.193 $\pm$0.034 &2MASS\\
$K_s$ (mag) & 6.68 $\pm$ 0.03 & 7.032 $\pm$0.063 &2MASS\\
$W1$ (mag) & 6.844 $\pm$ 0.060 &7.049 $\pm$0.081 &WISE \\ 
$W2$ (mag) & 6.748 $\pm$ 0.030 & 7.107$\pm$0.037&WISE\\
$W3$ (mag) & 6.777 $\pm$ 0.023 &7.056$\pm$0.029 &WISE \\
$W4$ (mag) & 6.668 $\pm$ 0.094 & 6.958$\pm$0.119 &WISE \\
\hline
\multicolumn{4}{c}{{\bf Kinematics }} \\
Barycentric RV (\kms) &  8.05$\pm$0.06  & 6.41$\pm$0.06 &  This paper \\
$U$ (\kms) & $ -8.71\pm0.04 $&$-9.27\pm0.04 $&This paper \\
$V$ (\kms) & $-21.50\pm 0.04$ &  $-20.28\pm0.04$ &This paper \\
$W$ (\kms) & $ -1.53\pm 0.04$ & $-0.47\pm0.04$ &This paper \\
\hline
\multicolumn{4}{c}{{\bf Physical Properties}} \\
Spectral type &   G6V$\pm$1  & K3V$\pm$1 &\citet{TorresSearch2006} \\
Rotation period (days) &   $2.85^{+0.04}_{-0.05}$ & unknown & This paper\\
\teff\ (K) & 5428 $\pm$ 80 & 4700$\pm$90 &This paper\\
\fbol\ ($10^{-8}$\,erg\,cm$^{-2}$\,s$^{-1}$) & 1.2026 $\pm$0.017 & 0.542 $\pm$ 0.008 & This paper \\ 
$M_*$ ($M_\odot$) &  1.01$\pm$0.06 & 0.84$\pm$0.06 & This paper \\ 
$R_*$ ($R_\odot$) & 0.964$\pm$0.029 & 0.864$\pm$0.036 &This paper \\
$L_*$ ($L_\odot$) & 0.725$\pm$0.013 & 0.327 $\pm$ 0.010 &This paper \\
Age (Myr) & 45$\pm$4 & 45$\pm$4 &\citet{Bell2015}\\
\vsini\ (km~s$^{-1}$) & 17.8$\pm$0.2  & 14.4$\pm$0.3 &This paper \\
$i_*$ (deg)\tablenotemark{a} & $> 82^{\circ}$ & \nodata &This paper \\
\enddata
\tablenotetext{a}{With the convention $i<90$.}
\end{deluxetable}

\section{Measurements}\label{Sec:measurements}

\subsection{Stellar parameters}\label{Sec:stellarparams}

{\it Age:} \system\ was one of the original systems used to define the Tuc-Hor moving group \citep[then called the Tucanae association,][]{ZuckermanWebb2000}. The group has consistent age estimates based on isochronal fitting \citep[45$\pm$4\,Myr; ][]{Bell2015} and the lithium-depletion boundary \citep[40\,Myr; ][]{Kraus2014}. Here we adopt the age estimate from \citet{Bell2015}. 

\begin{figure*}[ht]
    \centering
    \includegraphics[width=0.49\textwidth]{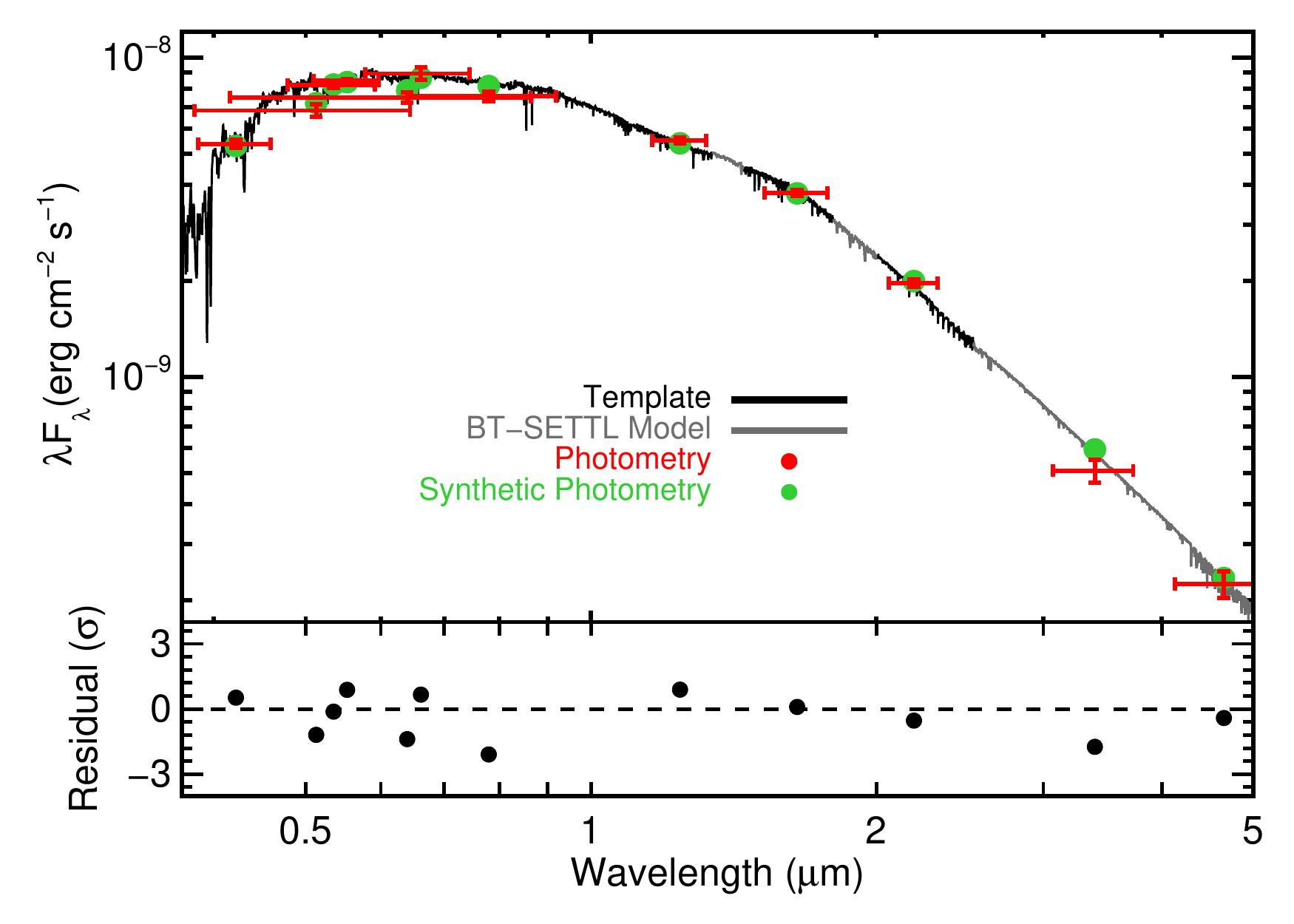}
    \includegraphics[width=0.49\textwidth]{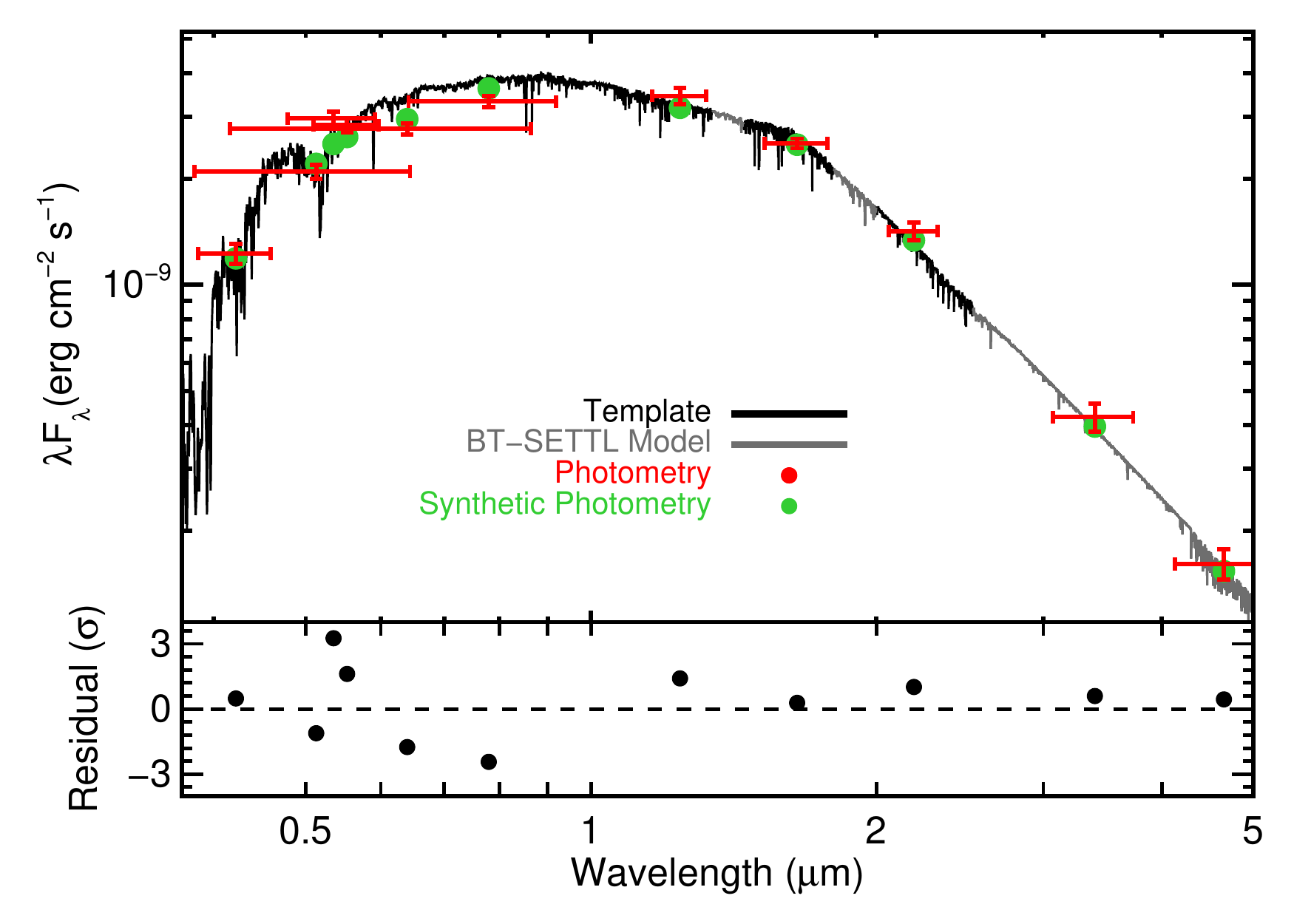}
    \caption{Best-fit spectral template compared to the photometry of \target\ (left) and \comp\ (right). Grey regions are BT-SETTL models, used to fill in gaps or regions of high telluric contamination. Literature photometry is shown in red, with horizontal errors corresponding to the filter width and vertical errors the measurement errors. Corresponding synthetic photometry is shown as green points. The bottom panel shows the residuals in terms of standard deviations from the fit.}
    \label{fig:sed}
\end{figure*}

{\it Luminosity, effective temperature, and Radius:} We first determined the bolometric flux (\fbol), \teff, and angular diameter of \target\ and \comp\ by fitting the resolved spectral energy distributions (SEDs) for each component with unreddened optical and near-infrared template spectra from the cool stars library \citep{Rayner2009}. A demonstration can be seen in Figure~\ref{fig:sed}. 

Our SED-fitting procedure followed the technique outlined in \citet{Mann2015b}, which we briefly summarize here. Our comparison assumed zero reddening, as \system\ lands within a region near the Sun of low interstellar extinction \citep[the Local Bubble;][]{LocalBubble}. We simultaneously compared each template spectrum to our optical spectra from SOAR/Goodman (Section~\ref{sec:goodman}) and archival photometry (Section~\ref{sec:phot} and Table~\ref{tab:sparams}) using the appropriate system zero-point and filter profile \cite{Cohen2003, Jarrett2011, Mann2015a, dr2_filter}. Gaps in each template spectrum are filled with a BT-SETTL atmospheric model \citep{Allard2012} using the model interpolation and fitting procedure described in \citet{Gaidos2014}. This procedure simultaneously provided an estimate of \teff\ based on the BT-SETTL model comparison to the observed spectrum. To compute \fbol, we integrated each template/model combination over all wavelengths. 

We combined the derived \fbol\ with the {\it Gaia} DR2 distance ($d$) to determine the total luminosity ($L_*$) for each component star. We then calculated a stellar radius ($R_*$) from $L_*$ and \teff\ using the Stefan-Boltzmann relation. Errors on each parameter were assigned accounting for both the measurement uncertainties (e.g., in the photometry) as well as the range of possible templates (and their assigned \teff\ values) that can fit the data. Final parameters and uncertainties are give in Table~\ref{tab:sparams}. 

As part of our above procedure, the BT-SETTL model is scaled to match the photometry and template. Assuming perfect models, this multiplicative scale factor is equal to $R_*^2/d^2$ \citep{Cushing2008}, which provided another estimate of $R_*$ given the {\it Gaia} DR2 distance. This technique is similar to the infrared-flux method \citep{Blackwell1977}. Radii derived from this scale factor are not totally independent of the above method, as they rely on the same photometry and models, but the latter technique is less sensitive to the assigned \teff. 

The first technique (Stefan-Boltzman) yielded a radius of 0.964$\pm$0.029$R_\odot$, and the scaling (infrared-flux method) yielded a consistent radius of 0.951$\pm$0.020$R_\odot$ for DS Tuc A. We adopt the former value for all analyses. 

{\it Mass:} We estimated the masses of \target\ and \comp\ by interpolating our luminosity estimates onto a modified isochrone grid from the Dartmouth Stellar Evolution Program \citep[DSEP, ][]{Dotter2008}. These grids were adjusted to include the effects of magnetic fields and where the boundary conditions are applied, as described in more detail in \citet{Muirhead2014}, \citet{Feiden2014a}, and \citet{Feiden2016}. We assumed solar metallicity, which is typical within a scatter of $\sim$0.1 dex for the young stellar populations in the Solar neighborhood (e.g., \citealt{2014A&A...568A...2S} and references therein). We used both 40\,Myr and 50\,Myr grids, using the spread to approximate errors introduced by the age uncertainty for the Tuc-Hor moving group. This interpolation yielded mass estimates of 1.01$\pm$0.06$M_\odot$ for \target\ and 0.84$\pm$0.06$M_\odot$ for \comp. We considered these errors to be slightly underestimated, as systematic differences between model grids can exceed 10\% at this age.

\vspace{1cm}
\subsection{Radial velocities}\label{sec:rvs}

We used high resolution data from HARPS, UVES, FEROS, SALT/HRS, and NRES/LCO to determine stellar radial velocities (RVs).
We measured RVs by computing the spectral line broadening function \citep[BF;][]{Rucinski1992} between DS Tuc A or B observations and a zero-velocity template. The BF represents the function that, when convolved with the template, returns the observed spectrum, carrying information on RV shifts and line broadening. Throughout the analysis we used the HARPS G2 binary mask as our template \citep[e.g.][]{Pepe2002}. A Gaussian profile was fit to the BF to determine the stellar RV. In each case the BF is single peaked and smooth, indicating a contribution from only one star. 

For each echelle order we computed a ``first pass'' BF, which was used to shift the observed spectrum near zero velocity. Orders that survive a 3$\sigma$-clipping algorithm were then stitched into three equal-length wavelength regions where the final BFs were computed. Our geocentric RV measurement and uncertainty were computed from the mean and standard deviation across these 3 regions. For archival observations that are provided as a single stitched spectrum, we created 150\,\AA \ wide initial ``orders''.

Finally, for each epoch we computed the BF for telluric absorption features using a continuum normalized A0 star as our template. These offsets were applied to our measured RVs. We have measured RVs for all archival data following the above procedure. While the HARPS pipeline provides more precise RVs, we preformed our own measurements to ensure the same zero-point corrections across different instruments. We found a $\sim$70 m s$^{-1}$ offset from the HARPS observations, similar to our measurement uncertainty, but recovered the same epoch-to-epoch variability.
Our final RVs are corrected for barycentric motion and listed in Table \ref{tab:rvs}.

As noted in the introduction, DS Tuc B was previously identified as a binary based on its RV variability and the presence of two spectral components. Our spectra are inconsistent with DS Tuc B having two near-equal spectral type components; for both stars at each epoch, there is only one peak in the BF. While the previous work did not give sufficient information to test the proposed scenario of RV variability, we also do not see evidence for RV variations in excess of reasonable jitter levels for young stars in either star.

\subsection{Projected rotation velocity}

\begin{figure}
    \centering
    \includegraphics[width=0.8\columnwidth]{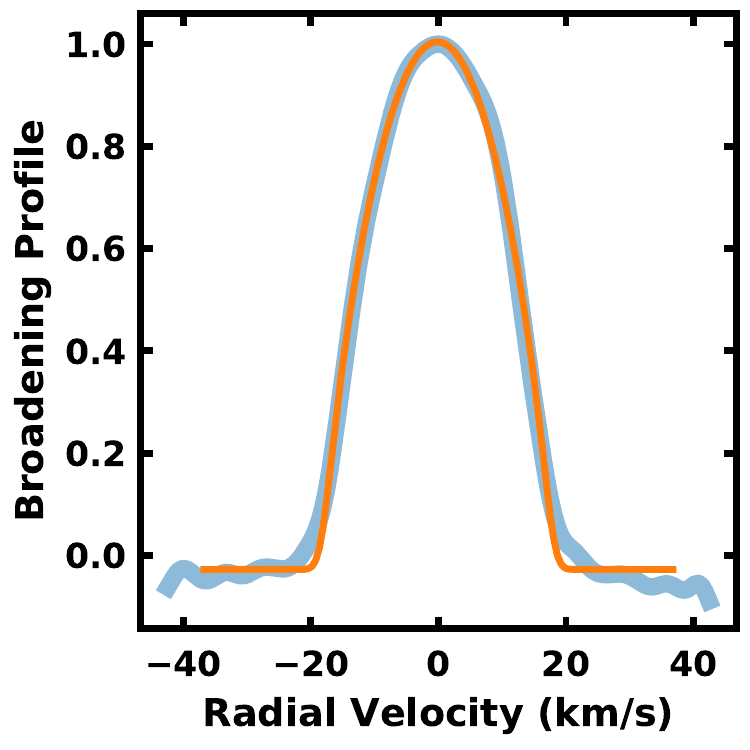}
    \caption{DS Tuc A broadening function computed from a representative HARPS spectrum. The broadening function presented in blue is clearly single-peaked and rotationally broadened. A best-fit rotational broadening profile is over plotted in orange. Extended wings in the broadening function as compared to the rotational broadening profile arise from additional line broadening mechanisms (macro/microturbulence) which are not included in our pure-rotation model. }
    \label{Fig:broad}
\end{figure}

We measured the projected rotational velocity (\vsini) for DS Tuc A and B by fitting the BF with a rotationally broadened absorption line profile that has been convolved with the instrumental profile (Figure \ref{Fig:broad}). We did not include additional broadening components such as microturbulence, though these factors should have minimal impact given the large \vsini values. For DS Tuc A, we find \vsini$=17.8\pm0.2$ km s$^{-1}$ using the HARPS spectra; the value is consistent when using SALT/HRS. From SALT/HRS observations of DS Tuc B, we measure \vsini$=14.4\pm0.3$ km s$^{-1}$.

\begin{deluxetable}{l r c c}
\tablecaption{Radial velocity measurements of DS Tuc A and B \label{tab:rvs}}
\tablewidth{0pt}
\tablehead{
\colhead{Site} & \colhead{BJD} & \colhead{RV} & \colhead{$\sigma_{RV}$}\\
\colhead{} & \colhead{} & \colhead{(km s$^{-1}$)} & \colhead{(km s$^{-1}$)}
}
\startdata
\multicolumn{4}{c}{{\bf DS Tuc A}} \\
HARPS & 2453500.876233 & 7.82 & 0.07 \\
HARPS & 2453521.828166 & 7.93 & 0.05 \\
HARPS & 2453522.888133 & 8.32 & 0.06 \\
HARPS & 2453541.927465 & 8.02 & 0.07 \\
HARPS & 2453600.704290  & 7.85 & 0.07 \\
UVES &  2454243.856154 & 8.27 & 0.10 \\
FEROS & 2455853.592265 & 7.98 & 0.24 \\
SALT &  2458439.283495 & 8.08 & 0.43 \\
SALT &  2458441.278033 & 8.29 & 0.46 \\
SALT &  2458442.295852 & 8.34 & 0.28 \\
SALT &  2458444.297823 & 7.74 & 0.31 \\
LCO &   2458463.540450  & 8.28 & 0.15 \\
\hline
\multicolumn{4}{l}{Mean: 8.05 (km/s)} \\
\multicolumn{4}{l}{RMS: 0.21 (km/s)} \\
\multicolumn{4}{l}{Std Error: 0.06 (km/s)} \\
\hline
\multicolumn{4}{c}{ {\bf DS Tuc B}} \\
SALT & 2458439.288665 &  6.41 & 0.31 \\
SALT & 2458441.273940  &  6.66 & 0.30 \\
SALT & 2458442.302087 &  6.42 & 0.21 \\
SALT & 2458444.302819 &  6.33 & 0.27 \\
UVES & 2454243.850252 &  6.25 & 0.11 \\
\hline
\multicolumn{4}{l}{Mean: 6.41 (km/s)} \\
\multicolumn{4}{l}{RMS:  0.14 (km/s)} \\
\multicolumn{4}{l}{Std Error:  0.06 (km/s)} \\
\enddata
\end{deluxetable}
\vspace{1cm}

\subsection{Stellar rotation}\label{sec:rotation}

{\it Rotation period:} A photometric rotation period of $2.85$ days for DS Tuc was previously reported by \citet{2012AcA....62...67K}, and is clearly visible in both the \tess\ and WASP lightcurves. Based on ground-based monitoring with the Las Cumbres Observatory, we associate this signal with \target. We break the WASP lightcurve into four $200$ day observing seasons and measure the rotation period and amplitude of variability in each season. The period is consistently $2.85$ days with high variability in the semi-amplitude ($2\%$ to $2.6\%$), but the phase shifts.
The periodogram shows power at the period and the first harmonic, and no additional signals are seen that could be associated with \comp.


The \tess\ lightcurve of DS Tuc shows consistent rotational modulation with a semi-amplitude of $1-2\%$. We modeled the \tess\ lightcurve with a Gaussian process (GP) using the \texttt{celerite} package from \citet{Foreman-MackeyFast2017}. We used a kernel composed of a mixture of simple harmonic oscillators and a jitter term. Our GP model has a term to capture the periodic brightness modulation caused by spots on the stellar surface. This kernel is a mixture of two stochastically-driven, damped harmonic oscillator models and has two modes in Fourier space: one at the rotation period of the star and one at half the rotation period. We initially included an additional damped harmonic oscillator with a period of $20$ days to capture long-term trends in the lightcurve, but the fitted power of the signal indicated that it was unnecessary.

We used a Lomb-Scargle periodogram to identify the candidate rotation period. We then fit the stellar rotation model using least squares, iterating 5 times and rejecting $3\sigma$ outliers each pass. This served to remove smaller flares. We then started an MCMC fit using the affine-invariant Markov Chain Monte Carlo (MCMC) implemented in the package \texttt{emcee} \citep{Foreman-MackeyEmcee2013}, beginning half the chains at the candidate rotation period identified in the periodogram, and a quarter each at half and twice the rotation period. We use 50 walkers and a burn-in of 5000 steps. We end the run when the autocorrelation timescale $\tau$ of all chains changes by $<0.1$ and the length of the chain is $>100\tau$. We measure a rotation period of $2.85^{+0.04}_{-0.05}$ days.

{\it Stellar inclination:} Following the method detailed in \citet{Morton2014b}, we combined the stellar rotation period measured from the \tess\ lightcurve, $R_*$, and \vsini\ measurements from above to estimate of the stellar inclination for \target. Although this measurement is not very precise, this method can identify highly misaligned systems \citep[e.g.,][]{Hirano_vsini2012} or be used for statistical studies of large planet populations \citep[e.g.,][]{2017AJ....154..270W}. We determine an equatorial velocity of 17.13$\pm$0.6\,km\,s$^{-1}$, consistent with our spectroscopic measurement of \vsini\ $=  17.8 \pm 0.2$\,km\,s$^{-1}$. This corresponds to a 1$\sigma$ lower limit on the inclination of $i > 82 ^\circ$ and a 2$\sigma$ lower limit of $i > 70 ^\circ$. We cannot distinguish between $i<90$\degree\ and $i>90$\degree, and so adopt the convention $i<90$\degree.

\section{Constraints on the DS Tuc system architecture}\label{sec:system}

\begin{figure*}
    \centering
    \includegraphics[width=0.5\textwidth]{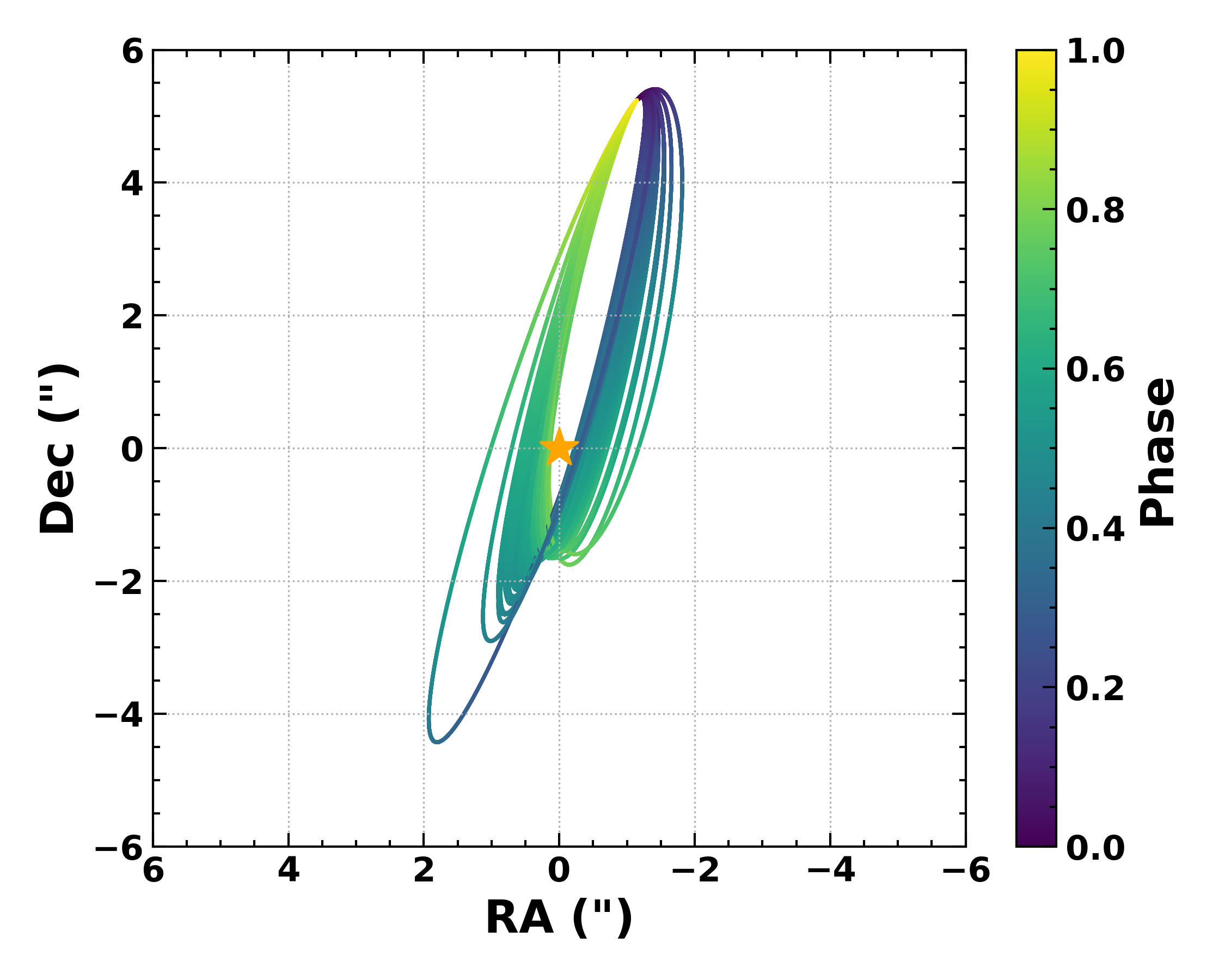}
    \includegraphics[width=0.96\textwidth]{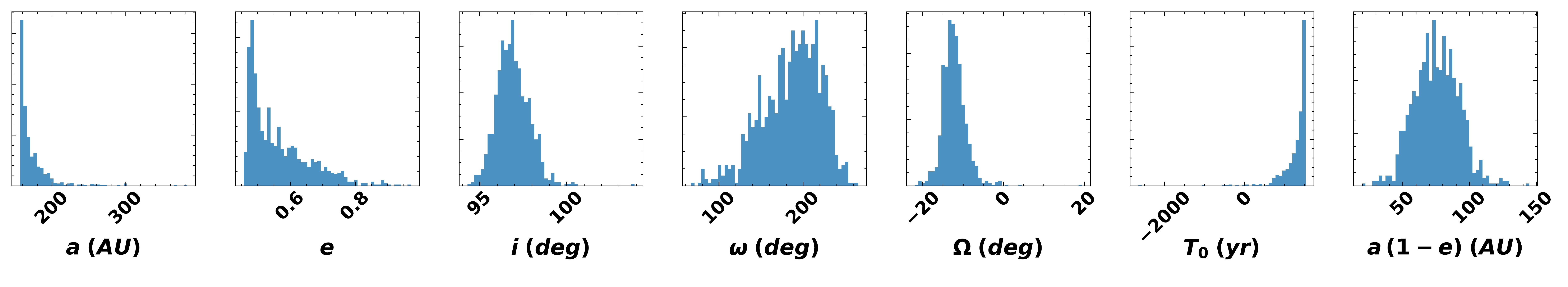}
    \caption{Top: 100 randomly selected orbits from the posterior distribution of accepted orbits for the stellar binary system.  \target\ is marked by the orange star at the origin, while the present position of \comp\ relative to A is located where the orbit tracks converge.  Orbital phase is shown by the color bar, with an orbital phase of 0.0 corresponding to the \textit{Gaia} observation epoch 2015.5.  
    Bottom: Posterior distributions for all orbital parameters from the fit, as well as periastron.  Semi-major axis and epoch of periastron passage have been truncated for clarity.  The inclination is tightly constrained to be nearly edge-on (90$^\circ$), close to the inclination of the transiting planet.}
    \label{fig:binaryorbits}
\end{figure*}

\begin{deluxetable*}{cccccc}[htb!]
\tablecaption{{Stellar Binary Orbital Parameters}\label{tab:binary orbits}}
\tablehead{\colhead{Element} &  \colhead{Median} & \colhead{Std Dev} &  \colhead{Mode} &
\colhead{68.3\% Min CI} & \colhead{95.4\% Min CI}  }
\startdata
$a$ (AU) & 176 & 29 & 160 & (157, 174) & (157, 219)\\
$P$ (yrs) & 1760 & 510 & 1500 & (1470,1730) & (1470,2440) \\
$e$ & 0.57 & 0.10 & 0.47 & (0.46, 0.60) &   (0.46, 0.77)\\
$i$ (\degree) & 96.9 & 0.9 & 96.6 & (96.0, 97.8) & (95.0, 98.6) \\
$\omega$ (\degree) & 186 & 35 & 196 & (164, 233) & (122, 256)\\
$\Omega$ (\degree) & -12 & 3 & -13 & (-15, -10) & (-18, -6)\\
$T_0$ (yr) & 1250 & 480 & 1520 & (1250, 1530) &  (-590, 1530)\\
Periastron (AU) & 75 & 17 & 85 & (59, 93) & (44, 105)\\
\enddata
\tablecomments{We report the median, mode, standard deviation, and 68.3\% and 95.4\% minimum credible intervals, with marginal posteriors and joint distributions displayed in Figure \ref{fig:binaryorbits}}
\end{deluxetable*}

\subsection{Stellar binary orbit}\label{sec:orbit}
We fit orbital parameters to the motion of the binary pair using a modified implementation of the Orbits for the Impatient (OFTI) rejection-sampling methodology described in \citet{Blunt2017}. This implementation is publicly available on GitHub\footnote{https://github.com/logan-pearce/LOFTI \citep{LOFTI}} and described further in \citet{Pearce2019}. 

Both objects have a well-defined \textit{Gaia} DR2 astrometric solution, so we used the positions and proper motions of \comp\ relative to \target\ in the plane of the sky.  We used the radial velocity measurements of Table \ref{tab:rvs} to interpolate a relative radial velocity at the \textit{Gaia} observation epoch of 2015.5.  Relative separation and position angle measurements in the Washington Double Star Catalog (WDS) spanning 126 years provide additional constraints on the stellar orbital motion.
We performed a modified OFTI fit constrained by these measurements.  

Previous implementations of OFTI have fit orbital parameters to astrometric observations spanning several epochs \citep[e.g.][]{Blunt2017, Pearce2019, Ruane2019, Cheetham2019}. In this system, the precision of the \textit{Gaia} solution for both objects allowed us to constrain five of the six position vector elements using just this single epoch, and we additionally have the astrometric measurements provided by WDS; only the line-of-sight position is not sufficiently constrained to contribute to the fit.

Table \ref{tab:binary orbits} displays the orbital parameters we determined for the stellar binary orbit.  Figure \ref{fig:binaryorbits} displays the orbital parameter distributions, joint credible intervals, and a selection of orbits plotted in the plane of the sky. The  orbital semi-major axis is $157 < a < 174$ au, with a closest approach of $59 < r_{peri} < 93$ au (where the ranges are $1\sigma$ credible intervals). The stellar binary is constrained to be nearly edge-on ($96.0^{\circ} < i < 97.8^{\circ}$), 
which is likely aligned with both the transiting planet's orbit and the primary star's spin axis.

\begin{figure*}[t]
    \centering
    \includegraphics[width=0.98\textwidth,trim=2cm 14cm 2cm 7.5cm]{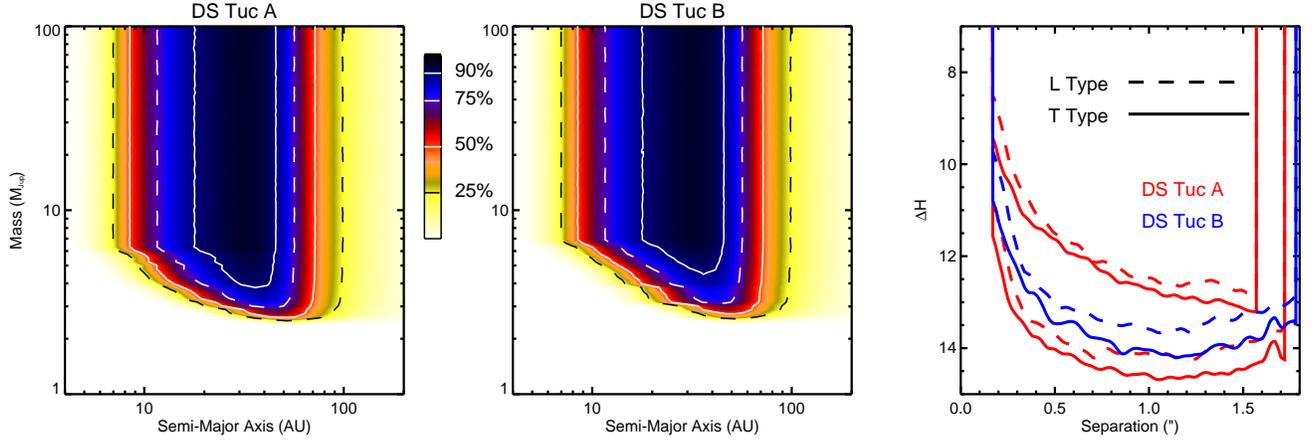}
    \caption{Left and Center: Completeness to substellar companions from the GPIES observations of DS Tuc A and B.  Planets and brown dwarfs more massive than $\sim$5 M$_{\rm Jup}$ are excluded at high completeness between 10--80 au.  Right: Contrast curves from which these completeness maps are derived, based on two epochs of GPIES observations of DS Tuc A, and one of B.  The contrast limits are slightly deeper for T-type spectra, as PSF subtraction can leverage the strong methane absorption for the coolest planets.}
    \label{fig:contrast}
\end{figure*}

\subsection{Limits on additional directly imaged companions}\label{sec:directimaging}

To search for companions in high contrast imaging data from GPI, we forward modeled the PSF template of a hypothetical companion at each pixel in the image using the Forward Model Matched Filter technique \citep[FMMF;][]{Ruffio2017a}. We then ran a matched filter with the template in an attempt to maximize the signal of a planet at that location in the image. The method accounts for the distortion of the signal due to the speckle subtraction step. The detection limits are expressed in terms of the flux ratio between the point source and the star and were calibrated using simulated point source injection and recovery. The detection limits are set at six times the standard deviation of the noise in the final image, which is calculated in concentric annuli as a function of separation to the star. This detection threshold ensures a false-positive rate of less than one per 20 sequence of observations. The default matched filter reduction used for GPIES assumes a featureless spectrum, corresponding to hot planets, for the estimation of the point-source brightness. However, \citet{Ruffio2017b} showed that it can be used for the detection of stars without loss of sensitivity. We did not detect any candidate companions above our detection threshold in either dataset.

We determined completeness to bound substellar companions using the method described in \citet{NielsenGPIES}.  An ensemble of simulated companions were generated with full orbital parameters at a grid of semi-major axis and planet mass.  The projected separation in arcseconds was then computed for each simulated companion given the distance to the star, and the contrast was calculated using the BT-Settl models \citep{Baraffeetal2015}, the age of the star (45 Myr), and the star's $H$ magnitude.  Each simulated companion was compared to the measured contrast curve, and companions lying above the curve were considered detectable.  The same simulated companions were compared to multiple contrast curves, advanced forward in their orbits when observations are made at different epochs, as is the case for DS~Tuc~A.  Outside a radius of $\sim$1.1\arcsec, not all position angles fall on the detector; to compensate, we reduce the completeness beyond $\sim$1.1\arcsec using the fractional coverage as a function of radius.  

The depth of search plots, giving completeness as a function of semi-major axis and companion mass, are given for DS Tuc A and B in Figure~\ref{fig:contrast}, along with the underlying contrast curves.  There are two contrast curves at each epoch, a T type curve assuming heavy methane absorption in the matched filter step (appropriate to companions as hot as $\sim$1100 K), and an L type contrast curve assuming a flatter spectrum appropriate to hotter brown dwarfs and stars.  Overall, wider separation planets and brown dwarfs are ruled out at high confidence between $\sim$10-80 au, more massive than $\sim$5 M$_{\rm Jup}$, around both A and B.

\subsection{Limits on wide binary companions}\label{sec:gaiawidecomps}

Past AO observations of the DS Tuc system have been limited to an outer working angle of $\rho \la 10$\arcsec\, \citep[e.g.][]{kasper2007}, leaving open the possibility of a hierarchical architecture with a very wide tertiary companion. The Gaia catalog reveals that there is one comoving, codistant candidate Tuc-Hor member within $<$1 pc of the DS Tuc system, 2MASS J23321028-6926537, which was also suggested to be a candidate low-mass (spectral type around M5) member of Tuc-Hor by \citet{2015ApJ...798...73G}. However, given the very wide separation ($\rho = 1.12 \times 10^5$ AU), this source is likely an unbound member of Tuc-Hor and not a bound companion of DS Tuc. There are no other candidate wide companions in Gaia DR2 within $\rho < 1$ pc and brighter than a limiting magnitude of $G \sim 20.5$ mag, corresponding to a mass limit of $M > 15 M_{\rm Jup}$ at $\tau = 40$ Myr \citep{Baraffeetal2015}.

\subsection{Limits on additional transiting planets}\label{sec:notchinjrec}

\begin{figure}
    \centering
    \includegraphics[width=\columnwidth]{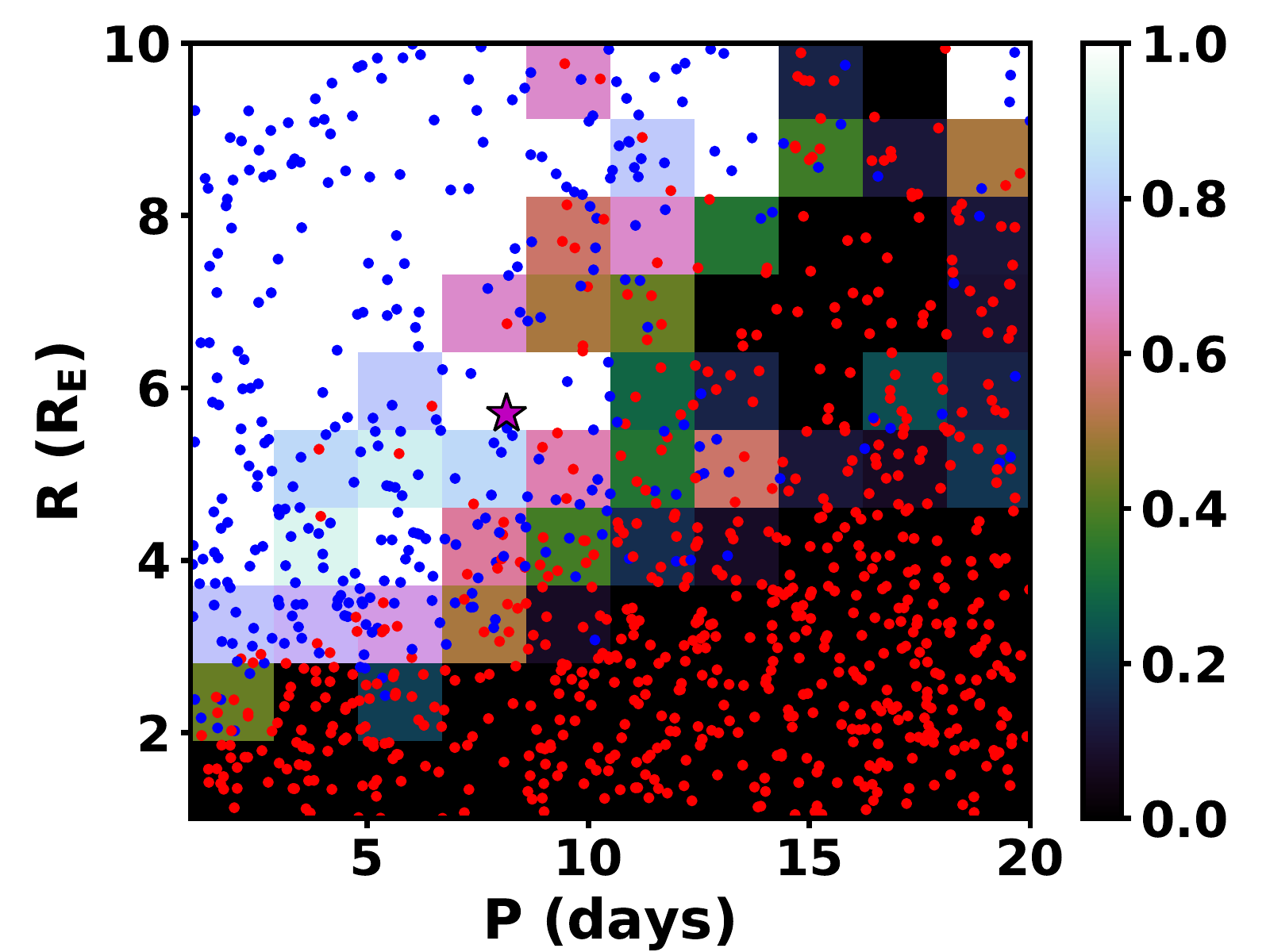}
    \caption{Completeness map for additional planets in the \target~ system, produced from injection-recovery testing of our search pipeline \citet{RizzutoZodiacal2017}. Each point represents an injected planet signal, with blue points indicating recovery and red points indicating non-recovery. The magenta star marks the position of the detected planet DS Tuc Ab.}
    \label{fig:transit_comp}
\end{figure}

We tested the detectability of additional planets in the \tess~sector 1 lightcurve of \target~using the notch-filter detrending and planet search pipeline of \citet{RizzutoZodiacal2017}. For this process, we used the SAP lightcurve which is not corrected for systematics using the cotrending basis vector method. This choice was made based on the presence of artifacts in the PDCSAP lightcurve, likely introduced by the presence of a strong stellar rotation signal. We first apply a deblending factor based on the \tess\ magnitudes for \target~and B and masked the time interval when fine-guiding was lost. We then injected a set of model transiting planets synthesized with the BATMAN model of \citet{KreidbergBatman2015} with orbital and size parameters chosen randomly. We used orbital periods of 1--20\,days and planet radii of 1--10\,$R_\earth$, and allowed orbital phase and impact parameter to take values in the interval [0,1]. Eccentricity was fixed to zero for this process, as it does not significantly influence detectability of a transit, but requires two additional variables over which to marginalize. We injected a total of 1000 trial planets for this test.

For each trial planet, we apply the notch filter detrending pipeline, and then search for periodic signals with the BLS algorithm \citep{kovacsBLS}, retaining signals with power-spectrum peaks above 7$\sigma$. We then set tolerance windows of 1\% in both injected period and orbital phase to flag a trial planet as recovered. Figure \ref{fig:transit_comp} shows the completeness map for additional planets in the \target~system. Our search and the \tess~sector 1 data for \target~are sensitive to $\sim$4\,$R_\earth$ planets at period $<$10\,days, and $\sim$3\,$R_\earth$ at periods $<$6\,days. At periods longer than 10 days, the time baseline and gaps due to the masked section significantly decrease sensitivity to transiting planets.

\section{Analysis of the planetary signal}\label{sec:planet}

\subsection{Identification of the stellar host}

The two components of DS Tuc are separated by $5\arcsec$\, and are not resolved by \tess,\footnote{The \tess\ alert somewhat arbitrarily identifies DS Tuc A as the host because it is the brightest star in the vicinity.} which has a plate scale of 21$\arcsec$ pixel$^{-1}$ with 50\% of light concentrated within one pixel \citep{2014SPIE.9143E..20R}. We examined the measured centroid of the in-transit/out-of-transit difference image, which is calculated by the SPOC pipeline and included in the data validation (DV) report (from the initial TPS run) that accompanied the alert. The DV report indicated that both DS Tuc A and B are contained within the 3$\sigma$ confusion radius of the centroid (which we note is dominated by the 2.5\arcsec additional error added in quadrature to the propagated uncertainty) and the centroid analysis averages a transit signal and a spurious event. In the second TPS run, not included in the alert, the centroid offset is consistent with DS Tuc A at 2$\sigma$.  We also analyzed the image centroids measured by the SPOC pipeline. The scatter in the centroid measurements is too large ($\simeq$ 1 millipixel per 4 hour bin) to detect the expected change in centroid position if the planet were to in fact orbit DS Tuc B (0.5 millipixel over a 3 hour transit). 
In summary, we found that the \tess\ data alone cannot conclusively identify which star hosts the transit. 

\begin{figure}
    \centering
    \includegraphics[width=\columnwidth]{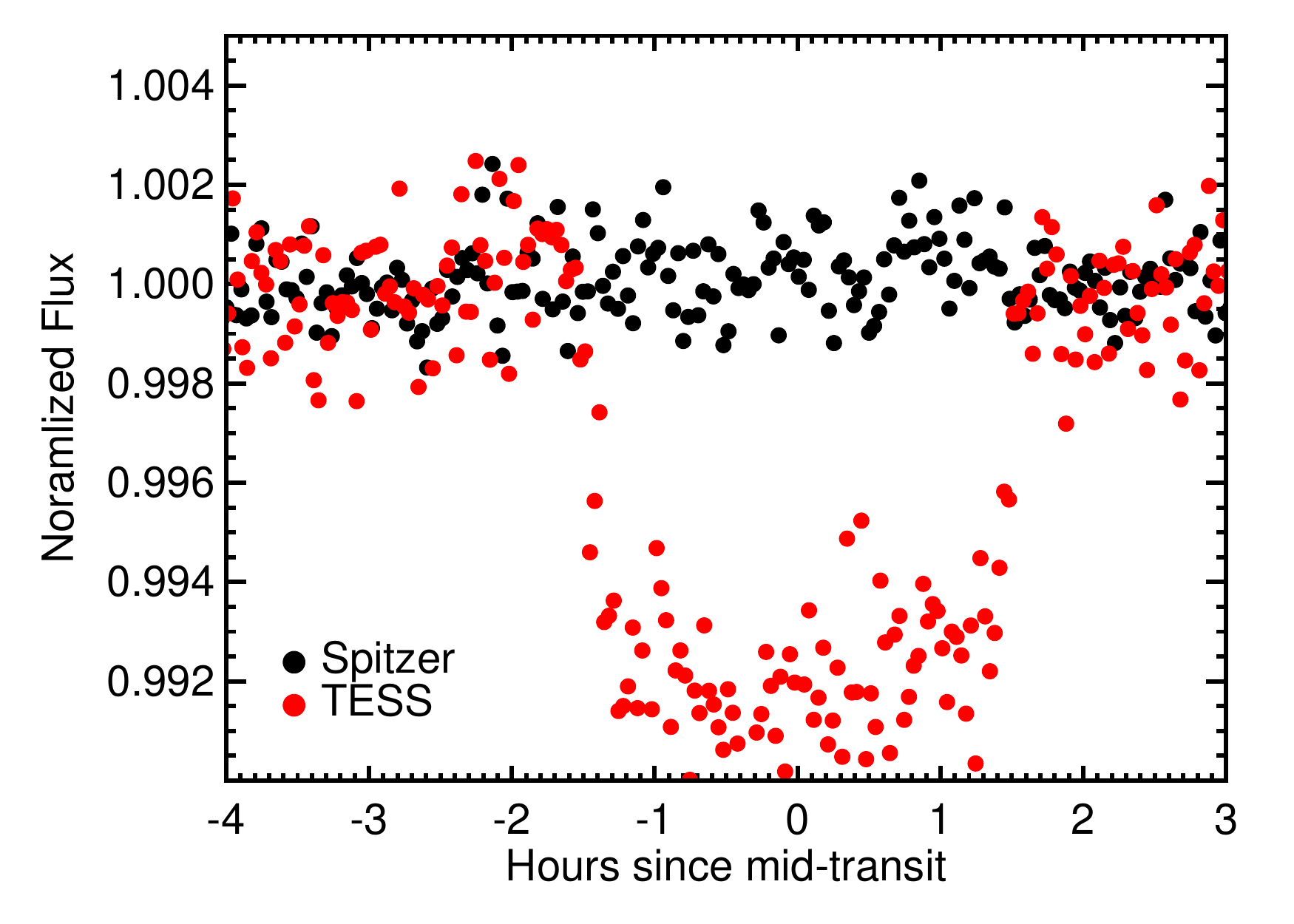}
    \caption{The {\it Spitzer} light curve from 2019 March 01 for a 4$\times$4 pixel aperture centered on \comp\ (black) compared to the {\it TESS} photometry at an aperture centered on \comp\ (red). The \tess\ data shown here assumes (incorrectly) that the planet orbits \comp, and it has been corrected for contamination from \target. Flux measurements from {\it Spitzer} were binned with 300 measurements per bin for clarity. In the resolved {\it Spitzer} data, \comp\ shows no transit signal and we thus conclude that the planet orbits DS Tuc A.}
    \label{fig:spitzer}
\end{figure}

Our \spitzer\ observations definitively show that the planet orbits DS Tuc A. A 4$\times$4 pixel aperture placed on DS Tuc A revealed a transit signal consistent with that detected in the \tess\ data. An equal-sized or smaller aperture centered on DS Tuc B yielded no detectable transit signature (Figure~\ref{fig:spitzer}).

\vspace{1cm}
\subsection{Transit fitting}\label{sec:transit}

We simultaneously fit the \tess\ and \textit{Spitzer} photometry using the transit fitting code \texttt{misttborn}.\footnote{\url{https://github.com/captain-exoplanet/misttborn}} \texttt{misttborn} was first used in \citet{MannZodiacal2016a} and has been used for a number of more recent works including \citet{JohnsonK22602018}. Briefly, we fit each system using \texttt{emcee}, and produced photometric transit models using \texttt{batman} \citep{KreidbergBatman2015}, which is based on the transit model of \citet{MandelAnalytic2002}. In the MCMC we fit for the following planetary parameters: the planet-to-star radius ratio ${R}_{P}/{R}_{\star }$ (assumed to be the same in all filters), impact parameter $b$, period $P$, and the epoch of the transit midpoint ${T}_{0}$. We fix eccentricity to zero. 
We also fit the following stellar parameters: linear and quadratic limb darkening parameters for each filter ($q_1,q_2$) using the triangular sampling method of \citet[][]{Kipping2013}, and the mean stellar density (${\rho }_{\star }$). We use Gaussian priors for the limb darkening parameters, using the values in \citet{ClaretAstronomy2011} and \citet{2017A&A...600A..30C}. We use uniform priors within physically-allowed boundaries for the remaining parameters (most notably, we enforced $|b| < 1+{R}_{P}/{R}_{\star }$ in order to assure that a transit occurs while allowing grazing transits).

DS Tuc is a visual binary with a separation of $\rho \sim 5 \arcsec$. The \tess\ photometry is de-blended, but the de-blending process may introduce errors, while our \spitzer\ aperture on \target\ includes a small amount of contamination from \comp. We included as an additional MCMC parameter the contamination of the aperture by flux from other stars. This is implemented as a (fractional) flux added to the transit model to create a diluted model ($LC_{\rm{diluted}}$) of the form;
\begin{equation}\label{eqn:dilution}
LC_{\rm{diluted}} = \frac{LC_{\rm{undiluted}}+C}{1+C},
\end{equation}
where $LC_{\rm{undiluted}}$ is the model light curve generated from {\tt Batman} and our GP model. This is comparable to the method used in \citet{2011ApJ...730...79J} and \citet{Gaidos2016b} to correct for flux dilution from a binary using the measured $\Delta m$ between components. The key difference is that Equation~\ref{eqn:dilution} allows for flux to be subtracted from the model ($C<0$) in the case of an over-correction. 

We set a Gaussian prior upon $C$ of $0.00\pm0.02$ for {\it TESS} and $0.0217\pm0.0050$ for {\it Spitzer}. The width of 0.02 for {\it TESS} photometry was estimated based on uncertainties in the derived \tess\ magnitudes from the TIC. Section~\ref{sec:spitzer} describes how $C$ for {\it Spitzer} was calculated from a model of the PSF. 

The target displays substantial stellar variability in the \tess\ bandpass. In addition to the transit model described above, we utilized Gaussian process regression to account for stellar variability in the \tess\ photometry. This enables us to model the variations in the stellar flux occurring during the transit. Our kernel is a mixture of simple harmonic oscillators, the same as described in Section~\ref{Sec:measurements}. We included the Gaussian process hyperparameters as fit parameters in our MCMC, and placed priors on those parameters based on the results of our stellar rotation modeling. The parameters are the stellar rotation period $P_*$, the amplitude $A_{\rm GP}$ of the primary signal at $P_*$, the relative strength of the secondary signal at $P_*/2$ (Mix$_{Q1,Q2}$), the decay timescales of the primary and secondary signals ($Q1_{\rm GP}$, $Q2_{\rm GP}$), and a jitter term to account for white noise ($\sigma_{\rm GP}$).\footnote{ https://celerite.readthedocs.io/en/stable/python/kernel/} 

We ran the MCMC chain with 100 walkers for 30,000 steps and cut off the first 5000 steps of burn-in, producing a total of 2.5$\times10^{6}$ samples from the posterior distributions of the fit parameters. The resulting fit is shown in Figure \ref{fig:transit}, and the best fitting values are listed in Table \ref{tab:params}.

\begin{deluxetable}{l c}
\tablecaption{Parameters of DS Tuc Ab \label{tab:params}}
\tablewidth{\columnwidth}
\tablehead{\colhead{Parameter} & \colhead{Value}}
\startdata
\multicolumn{2}{c}{{\bf Measured parameters}} \\
$T_0$ (TJD)\tablenotemark{a} & $1332.30997 \pm 0.00026$ \\ 
$P$ (days) & $8.138268 \pm 1.1\times10^{-5}$ \\
$R_P/R_{\star}$ & $0.05419 \pm 0.00024$ \\ 
$b$ & $0.18^{+0.13}_{-0.12}$ \\ 
$\rho_*$ ($\rho_\odot$) & $1.7^{+0.07}_{-0.17}$ \\ 
$q_{1,1}$ & $0.284^{+0.055}_{-0.053}$ \\ 
$q_{2,1}$ & $0.284 \pm 0.051$ \\ 
$q_{1,2}$ & $0.0266^{+0.0094}_{-0.0091}$ \\ 
$q_{2,2}$ & $0.054^{+0.014}_{-0.013}$ \\ 
$C_\mathrm{TESS}$ & $0.015^{+0.018}_{-0.017}$ \\ 
$C_\mathrm{Spitzer}$ & $0.0208^{+0.0049}_{-0.005}$ \\ 
$\ln{P_{\mathrm{*}}}$ (day) & $1.0606^{+0.0102}_{-0.0098}$ \\ 
$\ln{A_{\mathrm{GP}}}$ (\%$^2$) & $-10.87^{+0.11}_{-0.12}$ \\ 
$\ln{Q1_{\mathrm{GP}}}$ & $2.57^{+0.39}_{-0.37}$ \\ 
$\ln{Q2_{\mathrm{GP}}}$ & $0.052^{+0.027}_{-0.026}$ \\ 
Mix$_\mathrm{Q1,Q2}$ & $0.15^{+0.26}_{-0.11}$ \\ 
$\sigma_{\mathrm{GP}}$ & $-8.682 \pm 0.013$ \\ 
\hline 
\multicolumn{2}{c}{{\bf Derived parameters}} \\
$R_P$ ($R_\earth$) & $5.70\pm0.17$\\
$a/R_{\star}$ & $20.35^{+0.29}_{-0.69}$ \\ 
$i$ ($^{\circ}$) & $89.5^{+0.34}_{-0.41}$ \\ 
$\delta$ (\%) & $0.2936 \pm 0.0026$ \\ 
$T_{14}$ (days) & $0.13235^{+0.00049}_{-0.00039}$ \\ 
$T_{23}$ (days) & $0.11818^{+0.00039}_{-0.00057}$ \\ 
$T_{\mathrm{peri}}$ (TJD)\tablenotemark{a} & $1332.30997 \pm 0.00026$ \\ 
$g_{1,1}$ & $0.3^{+0.055}_{-0.054}$ \\ 
$g_{2,1}$ & $0.228^{+0.066}_{-0.06}$ \\ 
$g_{1,2}$ & $0.0172^{+0.0057}_{-0.0051}$ \\ 
$g_{2,2}$ & $0.145^{+0.024}_{-0.028}$ 
\enddata
\tablecomments{We report the median and 68\% confidence interval for each parameter. Associated probability distributions for key parameters are shown in Figure \ref{fig:transit}.}
\tablenotetext{a}{TJD is TESS Juldian Date, which is BJD$-2457000.0$}
\tablenotetext{b}{Although we allow $b$ to explore negative values, the absolute value of $b$ is listed since positive and negative values are degenerate. Similarly, we cannot distinguish between $i<90$\degree\ and $i>90$\degree\ and adopt the convention $i<90$\degree.}
\end{deluxetable}

\subsection{false-positive analysis}

Since we do not have dynamical (radial velocity) confirmation of DS~Tuc~Ab, we use our other observations to show that the transits are caused by a real transiting planet. We consider and rule out the following false-positive scenarios: 

\begin{enumerate}
    \item \textit{The transits are caused by instrumental artifacts or residuals from stellar variability:} Though there are only two transits in the \TESS\ dataset with amplitudes much lower than the amplitude of starspot variability, we confirm the transits with \Spitzer, conclusively ruling out an instrumental origin for the signal. The \Spitzer\ detection of the transits in the near infrared, at the predicted time and with the same depth as in \TESS\ rules out stellar variability as an origin, which should be significantly lower in the Spitzer bandpass and should not produce periodic transit-like signals. 
    \item \textit{DS Tuc A is an eclipsing binary:} Our radial velocity observations showed no variations large enough to be caused by a stellar companion. To test this, we generated 100,000 binaries with random (uniform) mass ratios, argument of periastron, phase, inclination, and eccentricty. The period was fixed at 8.138\,days, and inclination was restricted ensure the companion eclipses ($\gtrsim70$\degree). We then compared each synthetic binary's predicted velocities to the observed velocities assuming an extra jitter term in the velocities of 100\,m/s (from stellar variability). All generated binaries down to 20$M_J$ in mass were rejected at $>5\sigma$, and $>99\%$ were rejected down to 5$M_J$. 
    \item \textit{Light from a physically unassociated eclipsing binary star or transiting planet system is blended with light from DS Tuc:} Spitzer confirms that the transit signal detected towards DS Tuc A must originate from within a few arcseconds of the star. We detected no stars nearby DS Tuc in our GPI adaptive optics imaging, and other groups have previously detected no nearby stars in their own AO observations \citep{kasper2007, vogt2015}. Crucially, due to its proper motion, DS Tuc has moved over half an arcsecond with respect to stationary background sources between the different AO imaging epochs over the last decade, so we are able to definitively rule out background stars too close to DS Tuc A for GPI to resolve. 
    \item \textit{Light from a physically associated eclipsing binary or planet-hosting companion is blended with light from DS Tuc A:} For this to be true, DS Tuc A must have a binary companion close enough to escape detection by GPI (inside about 8 AU) and bright enough to cause the transit signal we see. The magnitude difference $\Delta m$ between DS Tuc A and the faintest companion which could contribute the transit signal is given by: 
    \begin{equation}
        \Delta m \lesssim 2.5 \log_{10}\left ( \frac{t_{12}^2}{t_{13}^2 \delta} \right )
    \end{equation}
    \noindent where $t_{12}$ is the duration of transit ingress/egress, $t_{13}$ is the transit duration from first contact (beginning of ingress) to third contact (beginning of egress), and $\delta$ is the observed transit depth \citep{vanderburg2019}. Fitting the \TESS\ light curve with MCMC, but without any constraints from the stellar parameters yields $\Delta m \lesssim 2.4$ (95\% confidence). From a 45 Myr MIST isochrone \citep{Dotter2016,Choietal2016} at solar metallically (provided in the \TESS\ bandpass), this magnitude difference corresponds to a companion star with a mass $>$0.63 $M_\odot$.
    
    To place a dynamical upper limit on the mass of a companion, we perform a Monte-Carlo simulation of companion orbits to DS Tuc A with randomly drawn isotropic inclinations, masses below 1$M_\odot$, and semi-major axes below 8 AU (holding the eccentricity to zero). For obits that produce semi-major amplitudes less than half the range of our RV observations (0.6 km s$^{-1}$), we find that we can exclude companion masses above 0.28 $M_\odot$ at 95\% confidence. The large discrepancy between these mass limits excludes this scenario at high confidence. 

\end{enumerate}

Our observational constraints confidently rule out these false-positive scenarios, so DS~Tuc~Ab is almost certainly a genuine exoplanet.

\section{Discussion}\label{Sec:discussion}

\subsection{DS Tuc Ab in context}

With an age of $\tau \sim 45$ Myr, DS Tuc Ab is one of the few transiting planets with ages $\tau < 100$ Myr, joining the planets K2-33b \citep{DavidNeptunesized2016, MannZodiacal2016}, V1298 Tau b \citep{DavidWarm2019} and AU Mic b (Plavchan et al.~submitted). At $V=8.5$, DS Tuc A is the brightest of these transiting planet host stars, closely followed by AU Mic at $V=8.6$. 

Using photometry from \tess\ and \spitzer, we determined that DS Tuc Ab has a radius of $5.70\pm0.17$ $R_\earth$, placing it in the sparsely populated realm of super-Neptunes and sub-Saturns. The planet is young enough that it likely still contracting due to internal cooling and may also be losing mass loss; models from \citet{2018ApJ...868..138B} suggest that its radius will shrink by $5-10$\% over the next few 100 Myr.

DS Tuc is a visual binary, and we find no evidence for additional massive companions in the system.  While DS Tuc B has previously been suggested to be a spectroscopic binary, we do not see two components in the spectrum of DS Tuc B at any observed epoch, a visual companion in high contrast imaging data, or periodic radial velocity variations at the precision of our data ($200$ m\,s$^{-1}$). The detection of planetary or substellar companions orbiting DS Tuc A exterior to DS Tuc Ab could indicate that dynamical interactions played a role in the present orbit of DS Tuc A; however, our high contrast imaging data from GPI shows no companions with masses more than about $5M_\mathrm{Jup}$ between $10$ and $80$ AU.

The orbit of the stellar binary is likely to be closely but not perfectly aligned with both the orbit of the transiting planet and the spin-axis of the planet-hosting star. We found a binary orbit inclination of $96.9\pm0.9$\degree, a planetary inclination of $89.5^{+0.34}_{-0.41}$\degree, and a stellar inclination of $i > 82^{\circ}$ ($1\sigma$ limit). The latter two quantities use the convention of $i<90$; however, $i>90$ is equally likely. Although the position angles are presently unconstrained, the chance of all three having the similar inclinations by chance is small, suggesting the three axes are in fact close to aligned. This is similar to the five-planet {\it Kepler}-444ABC system \citep{CampanteAncient2015a}. \citet{DupuyOrbital2016} found that the orbit of {\it Kepler}-444BC and the orbits of the planets around {\it Kepler}-444A have the same inclination angle, and suggested that the planets formed {\it in situ} in close orbits around {\it Kepler}-444A.

The stellar density that we determine from the transit fit differs from that which we calculate from the stellar parameters by $3\sigma$. The most likely reason is either errors in the model-derived stellar mass, or a mild eccentricity ($0.05 \lesssim e \lesssim 0.1$). While our mass estimate has formal errors of $\simeq$6\%, predictions from different model grids can vary by $\simeq$10\%. Moderate eccentricities have been found for some other young planets, including two in the Hyades \citep{2014ApJ...787...27Q, Thao2019}.

\subsection{Prospects for follow-up}

Due to the brightness of DS Tuc A, this system offers an exciting opportunity for detailed characterization of a young planet. Measuring the planetary mass would allow one to compare the planet's density to that of older planets. A distinct possibility is that mass estimates based on field-age planets represent an overestimate for DS Tuc Ab, given that the planet could still retain heat from its formation and might undergo future radius evolution as its atmosphere is sculpted by photoevaporative ultraviolet flux. While these processes would impact the planetary radius, they are not be expected to have a substantial impact on the planetary mass.

The \cite{ChenPROBABILISTIC2016} mass-radius relation, which are based on field-aged planetary systems, predicts a planetary mass of $28^{+35}_{-13}$ $M_{\oplus}$. The expected radial velocity (RV) semi-amplitude produced by DS Tuc Ab would then be $9^{+11}_{-4}$ ms$^{-1}$. As evidenced by the large error bars on the inferred planet mass, there are relatively few planets with sizes between Neptune and Saturn with measured masses; and the planetary mass--radius relation is poorly constrained for planets of this size. 

Measuring the Rossiter--McLaughlin effect would determine the sky-projected angle between the stellar rotational and planetary orbital angular momentum vectors, and test our hypothesis that the stellar spin and planetary orbital axes are aligned. We estimate the radial velocity amplitude due to the Rossiter--McLaughin effect using the relation $\Delta RV\simeq$0.65~\vsini$\left(\frac{R_P}{R_*}\right)^2\sqrt{1-b^2}$ \citep{2007ApJ...655..550G}, finding a predicted amplitude of 32\,m\,s$^{-1}$. 
Combining a spin-orbit misalignment measurement from Doppler Tomography \citep[e.g.,][]{Johnson2017} or the Rossiter--McLaughlin effect \citep[e.g.,][]{Narita_rossiter2010} with our measurement of $i_*$ from the rotation period and \vsini, one could measure full three-dimensional spin-orbit misalignment $\psi$. DS Tuc Ab joins a small number of planets where such measurements are possible.

Measuring RV signals on the scales noted above would be well within reach of current high precision RV instruments, but stellar activity poses a major challenge \citep[e.g.][]{SaarActivityRelated1997, PaulsonSearching2004}. DS Tuc A is a very magnetically active star, with $\log{R^\prime_{HK}=-4.09}$ \citep{HenrySurvey1996}. For stars like DS Tuc A, the stellar activity signal on many-day timescales (i.e., over many stellar rotation periods) is expected to be $100-200$ m/s based on the sample of active stars monitored with Keck by \citet{HillenbrandEmpirical2015}. While a jitter of this level would seem to preclude RV measurements of the planetary signal, stellar activity signals can be mitigated by simultaneously modeling the activity and planetary signals using, e.g.\ Gaussian processes, a process which would be aided by our knowledge of the star's photometric variability \citep[e.g.][]{HaywoodPlanets2014, RajpaulGaussian2015, 2016AJ....152..204L}. It is not clear how well the activity signal can be modelled and removed in an intensive RV campaign to measure a planet's mass or Rossiter--McLaughlin effect. 

We investigate prospects for atmospheric characterization with JWST by computing its transmission spectroscopy metric using Equation 1 of \cite{2018PASP..130k4401K}. We assume zero albedo and full day-night heat redistribution to estimate an equilibrium temperature for the planet of 850 K. We find a transmission spectroscopy metric is 264, which can be interpreted as the S/N with which its transmission spectrum is expected to be measured (assuming a cloud-free atmosphere) with a 10-hour observing program with the NIRISS instrument. This makes DS Tuc Ab an excellent target for observations with JWST. Finally, we note that it may be possible to detect the planetary exosphere, e.g. using He 10830\AA\ transit observations \citep{SpakeHelium2018,OklopcicNew2018}.

\section{Summary}\label{Sec:summary}

We report the discovery of a hot planet with a radius of $5.7\pm0.17R_\earth$ around the young star DS Tuc A (G6V, $V=8.5$) using data from NASA's \tess\ mission. The host star was one of the first identified members of the $45$ Myr old Tucana--Horologium association, and has a stellar companion orbiting at $157 < a < 174$ AU ($1\sigma$ interval). The \tess\ data alone were insufficient to validate the planet given the nearby stellar companion, so we used photometry from {\it Spitzer} to confirm that the planet orbits \target\ and revise the transit parameters. We find that the rotation axis of DS Tuc A, the orbital axis of the stellar binary, and the orbital axis of the planet are likely to be aligned. 

This $45$ Myr-old planet offers numerous opportunities for further characterization and illustrates the utility of \tess\ in furthering the study of planetary evolution.

\acknowledgements
The authors would like to thank R.~Angus, D.~Foreman-Mackey, and B.~Sowerwine for helpful conversations regarding this manuscript.
This work was supported by the \tess\ Guest Investigator program (Grant 80NSSC19K0636, awarded to AWM). 
ERN acknowledges support from the National Science Foundation Astronomy \& Astrophysics Postdoctoral Fellowship Program (Award \#1602597).
This work makes use of observations from the LCO network.
Based on observations obtained at the Southern Astrophysical Research (SOAR) telescope, which is a joint project of the Minist\'{e}rio da Ci\^{e}ncia, Tecnologia, Inova\c{c}\~{o}es e Comunica\c{c}\~{o}es (MCTIC) do Brasil, the U.S. National Optical Astronomy Observatory (NOAO), the University of North Carolina at Chapel Hill (UNC), and Michigan State University (MSU).
Some of the observations reported in this paper were obtained with the Southern African Large Telescope (SALT) through Dartmouth College.
This paper includes data collected by the \tess\ mission, which are publicly available from the Mikulski Archive for Space Telescopes (MAST). Funding for the \tess\ mission is provided by NASA's Science Mission directorate.
This research has made use of the Washington Double Star Catalog maintained at the U.S. Naval Observatory.
We would like to thank the University of North Carolina at Chapel Hill and the Research Computing group for providing computational resources (the Longleaf Cluster) and support that have contributed to these research results. 
We acknowledge the use of public TESS Alert data from pipelines at the TESS Science Office and at the TESS Science Processing Operations Center. Resources supporting this work were provided by the NASA High-End Computing (HEC) Program through the NASA Advanced Supercomputing (NAS) Division at Ames Research Center for the production of the SPOC data products. 

\vspace{5mm}
\facilities{TESS, SALT (HRS), SOAR (Goodman), WASP, Spitzer, LCO (NRES), CDS, MAST, Simbad}
\software{{\tt Astropy} \citep{2013A&A...558A..33A}, {\tt emcee} \citep{Foreman-MackeyEmcee2013}, {\tt celerite} \citep{Foreman-MackeyFast2017}}

\bibliography{ref_youngplanets}

\end{document}